\documentclass[lettersize,journal]{IEEEtran}
\usepackage{amsmath,amsfonts, amssymb}
\usepackage{siunitx}
\usepackage{algorithmic}
\usepackage{algorithm}
\usepackage{array}
\usepackage{adjustbox}
\usepackage{multirow}
\usepackage{pifont}
\usepackage{booktabs}
\usepackage{makecell}
\usepackage{textcomp}
\usepackage{stfloats}
\usepackage{url}
\usepackage{verbatim}
\usepackage{graphicx}
\usepackage{xcolor}
\usepackage{cite}
\usepackage{tikz}

\usepackage{tabularx}
\usepackage[hidelinks]{hyperref}
\usepackage{orcidlink}

\usepackage[acronym]{glossaries}

\newacronym{isa}{ISA}{Instruction Set Architecture}
\newacronym{dsp}{DSP}{Digital Signal Processing}
\newacronym{simd}{SIMD}{Single Instruction Multiple Data}
\newacronym{vr}{VR}{Virtual Reality}
\newacronym[firstplural=Processing Elements (PEs)]{pe}{PE}{Processing Element}
\newacronym{ar}{AR}{Augmented Reality}
\newacronym{xr}{XR}{Extended Reality}
\newacronym[plural=SoCs,firstplural=Systems-on-Chip (SoCs)]{soc}{SoC}{System-on-Chip}
\newacronym{roi}{ROI}{Region of Interest}
\newacronym{nvm}{NVM}{Non-volatile Memory}
\newacronym{imc}{IMC}{In-Memory Computing}
\newacronym{fc}{FC}{Fabric Controller}
\newacronym{mac}{MAC}{Multiply-Accumulate}
\newacronym{ml}{ML}{Machine Learning}
\newacronym{tcdm}{TCDM}{Tightly-coupled Data Memory}
\newacronym{ai}{AI}{Artificial Intelligence}

\newacronym{mram}{MRAM}{Magnetoresistive Random Access Memory}
\newacronym{sram}{SRAM}{Static Random Access Memory}
\newacronym{fpu}{FPU}{Floating Point Unit}
\newacronym{fsm}{FSM}{Finite State Machine}
\newacronym{cdc}{CDC}{Clock Domain Crossing}

\newacronym{hwpe}{HWPE}{Hardware Processing Engine}
\newacronym{dma}{DMA}{Direct Memory Access}
\newacronym{aimc}{AIMC}{Analog In-Memory Computing}
\newacronym{beol}{BEOL}{back end of line}
\newacronym{ip}{IP}{Intellectual Property}
\newacronym[firstplural=Standard Cell Memories (SCMs)]{scm}{SCM}{Standard Cell Memory}

\newacronym[firstplural=Deep Neural Networks (DNNs)]{dnn}{DNN}{Deep Neural Network}

\newacronym[firstplural=Convolutional Neural Networks (CNNs)]{cnn}{CNN}{Convolutional Neural Network}

\newacronym{hwc}{HWC}{Height-Width-Channel}
\newcommand{\modified}[1]{{#1}}
\newcommand{\hp}{{\textit{nominal}}}
\newcommand{\lp}{{\textit{low-power}}}
\newcommand{\mycircled}[1]{\raisebox{.5pt}{\textcircled{\raisebox{-.9pt} {#1}}}}

\begin{document}

\title{Siracusa: A 16 nm Heterogenous RISC-V SoC for Extended Reality with At-MRAM Neural Engine}

\author{Arpan~Suravi~Prasad\orcidlink{0009-0009-6031-6668},~\IEEEmembership{Graduate~Student~Member,~IEEE,}
        Moritz~Scherer\orcidlink{0000-0002-2762-2307},~\IEEEmembership{Graduate~Student~Member,~IEEE,}
        Francesco~Conti\orcidlink{0000-0002-7924-933X},~\IEEEmembership{Member,~IEEE,}
        Davide~Rossi\orcidlink{0000-0002-0651-5393},~\IEEEmembership{Member,~IEEE,}
        Alfio~Di~Mauro\orcidlink{0000-0001-6688-1603},~\IEEEmembership{Member,~IEEE,}
        Manuel~Eggimann\orcidlink{0000-0001-8395-7585},~\IEEEmembership{Member,~IEEE,}
        Jorge~Tomás~Gómez\orcidlink{0000-0001-7918-4655},
        Ziyun~Li\orcidlink{0000-0001-6070-6310},
        Syed~Shakib~Sarwar\orcidlink{0000-0002-0086-1076},
        Zhao~Wang,
        Barbara~De~Salvo\orcidlink{0000-0002-0810-9903},
        and Luca~Benini\orcidlink{0000-0001-8068-3806},~\IEEEmembership{Fellow,~IEEE}%
\thanks{Pre-print manuscript submitted for review to the IEEE Journal of Solid-State Circuits. This work was supported in part by Meta Reality Labs, in part by the KDT Joint Undertaking project TRISTAN under Grant 101095947, in part by the KDT Joint Undertaking project ISOLDE under Grant 101112274, in part by the Spoke 1 on Future HPC of the Italian Research Center on High-Performance Computing, Big Data and Quantum Computing (ICSC) funded by MUR Mission 4 - Next Generation EU, and in part by the Convolve project evaluated by the EU Horizon Europe under Grant 101070374 and supported by the Swiss State Secretariat for Education Research and Innovation under contract number 22.00150. \textit{(Corresponding author: Arpan Suravi Prasad.)}}%
\thanks{A. S. Prasad, M. Scherer, A. Di Mauro, and M. Eggimann are with the Integrated Systems Laboratory, ETH Z\"urich, 8092 Z\"urich, Switzerland; e-mail \{prasadar,scheremo,adimauro,meggimann\}@iis.ee.ethz.ch.}%
\thanks{F. Conti and D. Rossi are with the Department of Electrical, Electronic, and Information Engineering (DEI), University of Bologna, 40126 Bologna, Italy; e-mail: \{f.conti,davide.rossi\}@unibo.it.}%
\thanks{J. T. Gómez, Z. Li, S. S. Sarwar, Z. Wang, and B. De Salvo are with Meta Reality Labs, Burlingame, CA, 94010, USA; e-mail: \{jtgomez,liziyun,shakib7,zhaowang,barbarads\}@meta.com}%
\thanks{L. Benini is with the Integrated Systems Laboratory, ETH Z\"urich, 8092 Z\"urich, Switzerland, and also with the Department of Electrical, Electronic, and Information Engineering (DEI), University of Bologna, 40126 Bologna, Italy; e-mail lbenini@iis.ee.ethz.ch.}
}

\markboth{Journal of \LaTeX\ Class Files,~Vol.~14, No.~8, August~2021}%
{Shell \MakeLowercase{\textit{et al.}}: A Sample Article Using IEEEtran.cls for IEEE Journals}

\newcommand{\todo}[1]{\textcolor{red}{#1}}

\newcommand{\toreview}[1]{\textbf{\textcolor{green}{#1}}}

\newcommand{\done}[1]{\textcolor{black}{\textbf{#1}}}

\newcommand{\neureka}{N-EUREKA} 
\newcommand{\riscv}{RISC-V} 
\newcommand{\mcl}{Machine Learning} 
\newcommand{\cluster}{Cluster} 
\newcommand{\iodomain}{\textsc{IO-Domain}} 
\newcommand{\wmem}{\textsc{Weight Memory Subsystem}} 

\newcommand*\circled[1]{\tikz[baseline=(char.base)]{
            \node[shape=circle,draw,inner sep=0pt,fill=red, text=white] (char) {#1};}}

\maketitle

\begin{abstract}
Extended reality (XR) applications are Machine Learning (ML)-intensive, featuring deep neural networks (DNNs) with millions of weights, tightly latency-bound (10-20 ms end-to-end), and power-constrained (low tens of mW \modified{average power}).
While ML performance and efficiency can be achieved by introducing neural engines within low-power systems-on-chip (SoCs), system-level power for nontrivial DNNs depends strongly on the energy of non-volatile memory (NVM) access for network weights.
This work introduces \textit{Siracusa}, a near-sensor heterogeneous SoC for next-generation XR devices manufactured in 16 nm CMOS.
Siracusa couples an octa-core cluster of RISC-V digital signal processing cores with a novel tightly-coupled ``At-Memory'' integration between a state-of-the-art digital neural engine called \neureka{} and an on-chip NVM based on magnetoresistive memory (MRAM), achieving 1.7$\times$ higher throughput and 3$\times$ better energy efficiency than XR SoCs using NVM as background memory.
The fabricated SoC prototype achieves an area efficiency of 65.2 GOp/s/mm${}^\text{2}$ and a peak energy efficiency of 8.84 TOp/J for DNN inference while supporting complex, heterogeneous application workloads, which combine ML with conventional signal processing and control.
\end{abstract}

\section{Introduction}\label{sec:intro}
The effort to create an immersive approach for interacting with digital content is fueling
the growth of \gls{xr} applications in various sectors, including healthcare, education, and training. \gls{xr} applications can potentially revolutionize our approach to interact with digital content \cite{abrash_creating_2021} through seamless interaction with digital and real objects through context-aware processing, integrating natural user expressions with advanced algorithms.

Previous generations of \gls{vr} devices are bulky and offload most of their compute-intensive tasks to remote servers or personal gateways. In contrast, the next generation of \gls{xr} glasses will need to deliver competitive real-time performance in a lightweight, non-stigmatizing ``classic eyeglass'' form factor \cite{abrash_creating_2021}. 
However, there are several challenges to designing non-stigmatizing \gls{xr} glasses in addition to the miniaturization of electronics and tight integration of sensors and processors. Most importantly, \gls{xr} glasses must process sensor inputs in real-time to convert, among others, gestures, speech, and user gaze \cite{han_megatrack_2020, feng_real-time_2022, abrash_creating_2021} into actionable information. Furthermore, battery lifetime is a crucial design challenge; \gls{xr} glasses should last throughout a day of continuous use without requiring large and heavy battery packs.

A promising approach to overcome several key challenges of future \gls{xr} glasses is near-sensor or edge computing; processing data close to its source minimizes the latency between sampling and reacting to user input while reducing or eliminating the energy required to transfer high-bandwidth sensor data to remote servers, which represents a large share of power consumption in this class of devices \cite{abrash_creating_2021}. Additionally, privacy concerns are minimized by transmitting only highly compressed, non-personally identifiable sensor data from the near-sensor computing system.
Emerging solutions include innovations such as the integration of \gls{cnn} accelerators 3D-stacked directly on top of sensors, as demonstrated by Eki~\textit{et~al.}~\cite{eki_96_2021}.

While the promises of near-sensor computing are enticing, there are several practical challenges to leverage its advantages. A significant challenge of near-sensor \gls{xr} compute systems lies in balancing the performance and efficiency of classic computer vision workloads with modern \gls{dnn}-based data extraction. A common approach is to integrate core-based compute clusters with \gls{dnn} accelerators \cite{conti_marsellus_2023, di_mauro_kraken_2022, miro-panades_samurai_2023, yang_three-dimensional_2022, zhang_22nm_2022}. This approach enables high energy efficiency and throughput for critical tasks based on \glspl{dnn}, such as hand or eye tracking, while preserving general-purpose computing capabilities.
However, efficient collaboration of cores and accelerators is a complex design challenge. Unless cores and accelerators are tightly integrated through high-bandwidth shared memory, end-to-end performance suffers from added latency and energy consumption caused by extensive data movement between cores and accelerators. 

Another major challenge for edge computing systems in the \gls{xr} domain is efficient data movement, especially regarding loading neural network weights. Different strategies have been proposed to address this bottleneck, the most promising of which is the introduction of high-density \gls{nvm} on the chip, but conventional Flash \gls{nvm} cannot be integrated into scaled FinFET technologies.
Emerging \gls{mram} technology achieves comparable speed and read energy-per-access to \gls{sram} at significantly higher density and is compatible with \gls{beol} fabrication in FinFET processes~\cite{chiu_22nm_2023,lee_331_2023}.
As such, \gls{mram} opens up many integration opportunities in heterogeneous \glspl{soc}: emerging \gls{imc} such as in-MRAM \cite{chiu_22nm_2023} technology approaches merge \gls{nvm} with the accelerator, but must trade-off with storage density or compute precision, impacting the end-to-end performance of the system. On the other hand, integrating \glspl{nvm} as background (L2/L3) on-chip memory optimizes density at a significant price in access bandwidth, latency, and, ultimately, energy efficiency. 

In this work, we extend our ESSCIRC'23\cite {scherer_siracusa_2023} paper presenting \textit{Siracusa}, an all-digital \gls{soc} for \gls{xr} computing fabricated in TSMC\,16\,nm.
Siracusa introduces a novel architecture for tightly-coupled \gls{nvm} integration with our accelerator, N-EUREKA, enabling high access speed (9-cycle latency, 92 Gbit/s) without overheads in terms of storage density or readout efficiency.
Compared to \cite{scherer_siracusa_2023}, we focus this paper on the characterization of the on-chip \gls{mram} \gls{nvm}, which was not characterized in our previous work,  and on the acceleration architecture built around it.
We call our approach ``at'' \gls{mram} to distinguish it from ``in'' memory solutions, as it provides a new design point in the trade-off between accelerator energy efficiency and memory density. Our ``At-\gls{mram}'' approach differentiates itself from prior ``in-\gls{mram}'' solutions by enabling comprehensive system integration and a technology-agnostic perspective. This ensures its broad applicability in various platforms, requiring only an \gls{mram} \gls{ip} that is compatible with the selected technology node, thus avoiding the constraints and costs associated with technology-specific full-custom designs.

Furthermore, by coupling the At-\gls{mram} \neureka{} engine with a cluster of RISC-V cores via a low-latency interconnect and shared multi-banked \gls{sram}, our design enables efficient, zero-copy, collaborative execution of heterogeneous \gls{ml}, \gls{dsp}, and control workloads while maximizing end-to-end energy efficiency. 
Unlike other heterogeneous near-sensor SoCs \cite{greenwaves_technologies_gap9_nodate}, Siracusa is designed to meet the high frame rate ($> 30FPS$) and low power consumption (averaging in the lower tens of mW) required by edge visual processing tasks in \gls{xr} applications. These tasks frequently depend on \gls{dnn}-based feature extraction and \gls{dsp} pipeline. Close coupled integration of a powerful neural engine(10,368 1x8b multipliers), with a high bandwidth (\SI{92}{Gbit\per\second}), high capacity \SI{4}{\mebi\byte} \gls{mram} enables Siracusa to meet these critical latency and power constraints by avoiding costly off-chip data movement. 

Specifically, the contributions of this paper are as follows:
\begin{itemize}
    \item We introduce a novel At-MRAM computing architecture where \gls{nvm} is tightly coupled with compute engines that uses \SI{4}{\mebi\byte} of high-density, high-bandwidth, non-volatile on-chip \gls{mram} for \gls{dnn} weights, eliminating off-chip weight transfers for standard edge computing \glspl{dnn}. We demonstrate that our design reduces \gls{dnn} inference latency by 40\,\% and increases end-to-end energy efficiency by 3$\times$ compared to conventional (L3) \gls{nvm} integration schemes, which use \gls{nvm} as background memory.
    \item In addition to the non-volatile \gls{mram}, we integrate a \SI{4}{\mebi\byte} \gls{sram} tile memory, which minimizes feature map data movement between hierarchy levels. For networks larger than \SI{4}{\mebi\byte}, the proposed tile memory can be used as additional high bandwidth weight memory using a novel lightweight hardware page manager, allowing it to switch between weight pages seamlessly.
    \item We present a heterogeneous compute cluster architecture integrating At-\gls{mram} N-EUREKA, a cooperative, weight-precision-tunable Convolutional \gls{dnn} accelerator featuring a peak \gls{dnn} inference performance of \SI{1950}{GOp\per\second} with a state-of-the-art octa-core \riscv{} \cluster{} through a heterogeneous interconnect, enabling priority-based bandwidth sharing of the \cluster{}'s L1 memory.
\end{itemize}

The rest of this paper is organized as follows: in Section~\ref{sec:arch}, we discuss the \gls{soc} architecture of Siracusa. Section~\ref{sec:meas} presents measurement results and details Siracusa's performance characteristics on representative \gls{xr} workloads. Section~\ref{sec:casestudy} discusses the impact of tightly integrating \gls{nvm} with our accelerator and compares our design to conventional integration approaches. In Section~\ref{sec:discussion}, we compare our work to the previous state-of-the-art. Finally, Section~\ref{sec:conclusion} concludes this article, summarizing the results.

\section{Architecture}\label{sec:arch}

In this section, we introduce the hardware architecture of Siracusa. Furthermore, we demonstrate how the tight coupling of \gls{nvm} with the system's main \gls{dnn} compute engine enables ultra-efficient zero-off-chip transfer inference of state-of-the-art \gls{xr} workloads.

\begin{figure*}
  \begin{center}
\includegraphics[width=\linewidth]{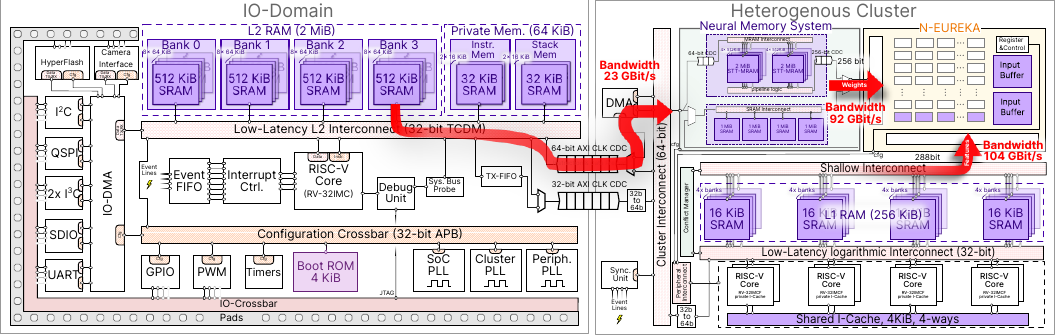}
\caption{Architectural overview of the Siracusa SoC consisting of the IO-Domain (upper left), a Heterogenous Cluster which includes the bit-serial N-EUREKA hardware accelerator (upper right) and 8 RISC-V cores (lower right). N-EUREKA is tightly coupled to the neural memory subsystem consisting of \SI{4}{\mebi\byte} SRAM and \SI{4}{\mebi\byte} of proprietary STT-\gls{mram} IP. The \riscv{} cores and \neureka{} share access to the L1 memory (middle right) through the heterogeneous interconnect, consisting of a shallow and logarithmic branch, synchronized by a conflict manager with programmable access priority.}
\label{fig:siracusa_block_diagram}
\end{center}
\end{figure*}

The Siracusa \gls{soc} comprises two domains: The \iodomain{} and the Heterogeneous \cluster{} as shown in Figure \ref{fig:siracusa_block_diagram}. The \iodomain{} consists of an advanced microcontroller based on a 32-bit \riscv{} \gls{fc} core responsible for management tasks, using the \texttt{RV32IMC} instruction set with \texttt{Xpulp} \gls{isa} extensions. The \texttt{Xpulp} extensions enhance throughput with zero-overhead hardware loops, post-increment load-store instruction, fixed-point operations, bit manipulations, and dot product instructions for 16-bit and 8-bit data. 

The \iodomain{} includes a large \SI{2}{\mebi\byte} L2 \gls{sram} scratchpad memory divided into four \SI{512}{\kibi\byte} word-interleaved banks. An additional \SI{32}{\kibi\byte} memory is reserved for the \gls{fc} program code and stack. The boot code for the core is stored in a dedicated \SI{4}{\kibi\byte} boot ROM. 
The \iodomain{} in Siracusa, while not the focus of our paper, serves essential roles, supporting diverse peripherals for real-world applications, such as streaming and storing images via the I3C interface for processing. Its inclusion underlines the chip's broader utility beyond the computational analysis emphasized in this paper.

The \iodomain{} incorporates an IO-\gls{dma} controller to transfer data between the L2 memory and IO interfaces efficiently. The supported IO interfaces include SPI, I3C, HyperRAM, and UART, as well as a dedicated camera interface with the image sensor to be tightly integrated for near-sensor computing \cite{abrash_creating_2021}\cite{murakami_49mpixel_2022}. This rich set of IO interfaces enhances Siracusa's adaptability and versatility to a diverse group of external devices, thus enabling its usability in various applications. The \iodomain{} contains three individual PLLs to generate clocks for the \iodomain{}, \cluster{} and the IO-Peripherals. The \gls{fc} accesses the PLLs, GPIOs, Debug Unit, and timers via a 32-bit APB crossbar. The \cluster{} is located in a separate power and frequency domain, consisting of 8 \riscv{} cores, a \gls{cnn} accelerator, \neureka{}, and a tightly coupled Neural Memory Subsystem. The \iodomain{} and the \cluster{} are connected via a \gls{cdc} into the 64-bit AXI \cluster{} interconnect. 

\subsection{Heterogeneous Cluster}\label{sec:archi_rv_cluster}
The Heterogeneous \cluster{} is an \gls{ai} and \gls{dsp} accelerator organized around two cooperative engines: a parallel set of 8 \riscv{} cores tailored to \gls{dsp} tasks, and a specialized hardware engine for quantized \gls{dnn} inference, called \textit{N-EUREKA}.

The 8 \riscv{} cores utilize the \texttt{RV32IMCF} \gls{isa}, similar to the \gls{fc}.
Furthermore, the cores are enhanced with the \texttt{Xpulpnn} instruction set, a superset of \texttt{Xpulp} adding support for dot-product for sub-byte 2, 4, 8-bit data and enhanced MAC\&LOAD instructions overlapping data loading with computation \cite{conti_marsellus_2023}.
The \riscv{} cores achieve near-100\% \gls{mac} utilization for linear algebra-dominated kernels, such as matrix multiplications and linear transformations.
The cores also support floating-point instructions via a dedicated \gls{fpu}.
In addition to the single-precision standard extension \texttt{F}, the floating point units support short-float formats such as FP8, FP16, FP16Brain~\cite{montagna_low-power_2022}, catering to the needs of a diverse range of visual processing applications with different properties in terms of quantization noise resilience; in Siracusa, we employed a configuration with one private floating point unit per core using two pipeline stages. A detailed discussion of the compute cluster design and performance measurements may be found in Montagna et al. \cite{montagna_low-power_2022}.

The \riscv{} cores are equipped with a hierarchical instruction cache (I\$) implemented with \SI{4}{\kibi\byte}, 4-way associative, \SI{128}{\bit\per line} shared among all 8 cores and one dedicated \SI{512}{\byte} 4-way associative cache per core \cite{jie_energy-efficient_2020}. The I\$ is implemented with \glspl{scm} to enable multiport access efficiently compared to regular \gls{sram} cuts. 

Along with the \riscv{} cores, the \cluster{} domain also includes \neureka{}, a configurable hardware accelerator designed to perform \gls{dnn} inference tasks efficiently. \neureka{} supports 3$\times$3 dense, 3$\times$3 depthwise, and 1$\times$1 dense convolutions with 8-bit activation and configurable weight precision from 2-8 bits.
The internal architecture of \neureka{} is described in greater detail in Section~\ref{sec:archi_neureka}.
Alongside the two main compute engines, the \cluster{} includes many auxiliary peripherals, such as a \gls{dma} controller and an event unit, which helps orchestrate high-performance parallel execution of the \riscv{} cores and concurrent DMA transfers. The \riscv{} cores configure the DMA, Event Unit, and \neureka{} via a 32-bit peripheral interconnect. 

To support the execution of cooperative tasks, the \riscv{} cores and \neureka{} share a single \SI{256}{\kibi\byte} L1 \gls{tcdm} organized into 16 word-interleaved banks.
The \gls{tcdm} is accessible via a low-latency L1 interconnect, enabling access to the \gls{tcdm} at an aggregate bandwidth of up to \SI{184}{\giga\bit/\second} at \SI{360}{\mega\hertz}. This is achieved using 16 banks, each offering a bandwidth of 32 bits per cycle.
In the absence of bank contention, accesses are performed in a single cycle of latency.
The shared nature of L1 memory access mitigates superfluous data movements and enables pipelined, collaborative execution of tasks on \neureka{} and the \riscv{} cores, maximizing performance and efficiency.

The low latency L1 interconnect is organized hierarchically into a logarithmic and a shallow branch.
The logarithmic branch is used for arbitration between the 8 \riscv{} cores, the \gls{dma}, and as an external port towards the \iodomain{}.
The accesses are routed to each memory bank; the logarithmic branch employs a bank-level round-robin arbitration scheme in the event of memory contentions, thus avoiding core starvation.
In contrast, the shallow branch from \neureka{} routes accesses from a single contiguous-by-construction high bandwidth (\SI{288}{\bit\per cycle}) \neureka{} port towards the L1 memory banks, without any bank-wise arbitration.
The accesses of each of the two branches to the \gls{sram} \gls{tcdm} banks are further arbitrated using a configurable priority arbitrator without starvation, which can guarantee a minimum share of bandwidth for either branch while maximizing access bandwidth for the other~\cite{prasad_specialization_2023}.

\subsection{Neural Memory Subsystem}
\label{sec:archi_wmem}

In addition to L1 \gls{tcdm} memory, the Siracusa \cluster{} also introduces a dedicated Neural Memory Subsystem to enable the At-Memory acceleration concept discussed in Section~\ref{sec:intro}.

\subsubsection{Weight MRAM and tile SRAM subsystems}
The Neural Memory Subsystem is synchronous at the interface to the rest of the \cluster{} and comprises two distinct \SI{4}{\mebi\byte} subsystems.
The first subsystem is a novel \textit{weight memory subsystem} implemented using non-volatile \gls{mram}.
It is exclusively dedicated to storing \gls{dnn} weights, which are accessible to the \neureka{} accelerator contention-free and at high bandwidth, with a tight integration scheme detailed in Section~\ref{sec:archi_neureka}.
The \gls{mram} is organized internally in four banks of \SI{1}{\mebi\byte} each, which reside in a dedicated frequency/voltage domain, isochronous to the rest of the \cluster{} through a clock divider defaulting to 1/2 setting.
Each bank is divided into two 512 KiB cuts accessible in parallel (for a total bandwidth of up to 512 bits/cycle @ 180 MHz) to provide the required bandwidth for \neureka{} while coping with the divided clock.

The second subsystem in the Neural Memory Subsystem is the \textit{tile memory subsystem}, implemented using conventional \gls{sram}.
The primary role of this volatile memory is to store intermediate feature map tiles without relying on the \iodomain{} L2 and using the \iodomain{}-\cluster{} \gls{cdc}.
As a secondary purpose, this memory could also be used for additional (temporary) storage of weights streamed from off-chip Flash for \glspl{dnn} with a larger memory footprint than the available on-chip \gls{mram}, to allow scalability of the workload at the cost of performance, or for weights that are frequently field-tuned during system operation (e.g., for continuous learning applications \cite{rusci_-device_2023}).

Both subsystems share a single 64-bit AXI read and write port to the \cluster{} interconnect (up to \SI{23}{\giga\bit/\second} @ \SI{360}{\mega\hertz}) used to access the Neural Memory Subsystem from the \cluster{} through the AXI interconnect for \gls{mram} configuration, initial weight programming, activation load/store access, and general use from the \cluster{}'s cores.
A wide (256-bit, up to \SI{92}{\giga\bit/\second} @ \SI{360}{\mega\hertz}) read-only port provides the At-Memory access from the accelerator at high bandwidth and contention-free, with only nine cycles of access latency.
Section~\ref{sec:archi_neureka_wmem} discusses the pipelining scheme employed by \neureka{} to access the weight memory subsystem at high bandwidth, while an in-depth discussion of the weight memory subsystem, comparing it to several state-of-the-art alternatives, is included in Section~\ref{sec:casestudy}.

\subsubsection{Software-assisted virtual memory paging}
Optionally, the Neural Nemory Subsystem can operate in a lightweight, software-assisted virtual memory mode where \neureka{} operates on virtual \qty{4}{\mebi\byte} pages.
This mode provides complete functionality when running larger networks, which require more than \SI{8}{\mebi\byte}.
In this mode, Siracusa must use an off-chip background memory and use the system \iodomain{} L2 memory to hold intermediate activations.
In the virtual memory mode, the Neural Memory Subsystem is used as a cache for memory pages of the \gls{dnn} weights allocated in off-chip Flash.
A small page handling circuitry maps \neureka{}'s transaction to either of the two physical memory pages (residing in the tile \gls{sram} and weight \gls{mram}, respectively) by comparing the address prefix with the two live page index registers that are exposed via the Neural Memory Subsystem's configuration interface.

Unless a transaction's page index matches either index register, it is stalled, and a page-miss interrupt is raised towards the \gls{fc}, which programs the IO-DMA to perform a weight page swap through the 32-bit AXI \gls{cdc}.
This happens concurrently with L2-L1 DMA transfers, which can support tiling of activations via the separate 64-bit \gls{cdc} port.
Once finished, the \gls{fc} updates the page index register, unblocking the stalled transaction, which is completely transparent to N-EUREKA.
The system also supports proactive page swapping that takes advantage of the typically deterministic weight access pattern in \gls{dnn} workloads.
Such a swapping can be triggered by the \gls{fc} on a page-switch interrupt, which allows for transparent network reconfiguration with negligible increase in overall circuit area and minimizes stall time.
\subsection{\neureka}\label{sec:archi_neureka}

Given the pervasiveness of \glspl{dnn} in modern \gls{xr} pipelines \cite{abrash_creating_2021, yang_three-dimensional_2022}, one of the design goals of Siracusa is state-of-the-art energy-efficient neural network inference acceleration. To effectively exploit the advantages of modern quantized neural networks, \neureka{} supports integer mixed-precision inference. This Section introduces the working principle and architecture of \neureka{} and highlights the close integration of the \gls{mram} weight memory. 

\subsubsection{Operating principle}
\neureka{} is a programmable hardware accelerator that supports 3$\times$3 dense, 3$\times$3 depthwise, and 1$\times$1 dense convolutions. It supports 8-bit input activation, 2-8-bit weight precision, and 8-bit requantized or 32-bit outputs. The  focus on low-precision quantization of weights over that of activations in \neureka{} is driven by the simpler mapping of different operators on the same hardware, reduced data movement, and smaller weight memory footprint, making it feasible for more networks to achieve fully on-chip processing reaping the benefits of the at-MRAM computing model.
We contrast this choice with fully mixed-precision approaches like Marsellus\cite{conti_marsellus_2023}, which offers full flexibility at a cost to performance and energy efficiency.
\neureka{}'s datapath is designed for bit-serial arithmetic, performing higher bit-width product calculations using bit shifting and repeated addition. \neureka{} performs 3$\times$3 dense and depthwise convolution in a bit serial manner, whereas it exploits the adder and shifter to perform 1$\times$1 dense convolution in a bit parallel mode. Requantization is achieved by an integer-domain affine projection: \neureka{} uses per-channel scaling, bias, and shift parameters, which are applied according to the requantization scheme proposed by Conti \cite{conti_technical_2020}.

\begin{figure*}
    \centering
    \includegraphics[width=0.97\textwidth]{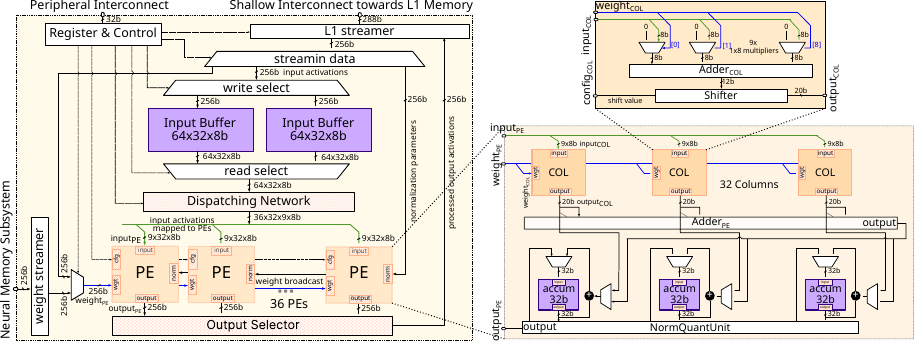}
    \caption{Overview of the datapath architecture of \neureka. The core of \neureka's datapath consists of 36 \glspl{pe}, which receive input activations from dual inputs buffers and weights from a dedicated weight streamer. The L1 streamer feeds into the input buffers and transfers outputs to the shared L1 memory. A detailed overview of the \gls{pe} datapath is shown on the right. Each \gls{pe} contains 32 columns, each containing nine bit-serial multipliers, an adder, and a shifter. Each column is connected to a dedicated \gls{scm} accumulator used to store partial results.} 
    \label{fig:neureka_archi}
\end{figure*}
\subsubsection{Microarchitecture}
Figure~\ref{fig:neureka_archi} shows the architecture of \neureka{}. We divide the architecture into three main components: two \textsc{streamers}, a \textsc{controller}, and a \textsc{datapath} unit. 
The \textsc{controller} consists of a latch-based dual context register file programmable by any \riscv{} core in Siracusa through the peripheral interconnect.
\neureka{} hosts two streamers: an L1 streamer, which provides 288-bit load / store access to L1 memory via the \gls{tcdm} interconnect, and a weight streamer, toward the Neural Memory Subsystem.
The L1 streamer is time division multiplexed to load input activations, normalization parameters, and optionally weights and to store processed output activation. 
Both streamers internally support the same bandwidth of \SI{256}{\bit} of contiguous data per cycle in either direction.
The available bandwidth is constrained by the available \gls{mram} bandwidth as explained in Section~\ref{sec:archi_neureka_wmem}; the same bandwidth is used in the shallow L1 interconnect to simplify the accelerator's control.
To handle the case of data that are not aligned on a word boundary in word-interleaved memory, \neureka{}'s L1 memory interface can load or store up to \SI{288}{\bit\per cycle} of data from contiguous banks.
The accelerator's L1 streamer automatically selects the relevant 256 bits from this larger block.
Fetching an extra word to compensate for unaligned accesses is necessary for L1 data, as they could be produced by the RISC-V cores; it is not required for the weight streamer, which accesses weight data that is aligned by construction.
Equipped with support for three-dimensional strided address generation, both streamers can serialize any 3D data access pattern and convert them to memory access transactions.

\neureka{}'s datapath consists of \gls{scm}-based \textsc{InputBuffer}s to store 8$\times$8$\times$32 input activation tiles and 6$\times$6 \glspl{pe}, allowing the accelerator to process 8$\times$8 spatial dimension tiles. Each \gls{pe} consists of 288 1$\times$8 bit multipliers organized into 9 rows and 32 columns. Each \gls{pe} also has 32$\times$32-bit accumulators to store partial sums. One additional \textsc{NormQuantUnit} per \gls{pe} performs scaling, adding a bias, and right-shifting, which are used to requantize higher bit-width outputs to 8 bits.

\subsubsection{Execution}
\begin{figure*}
    \centering
    \includegraphics[width=0.97\textwidth]{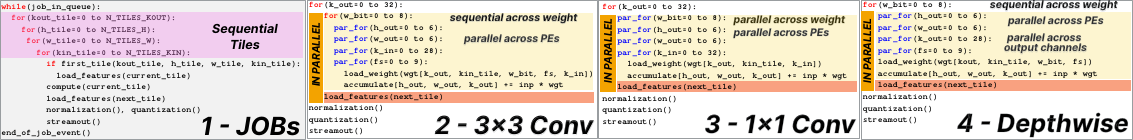}
    \caption{\modified{Overview of execution of a single layer on \neureka{}. \mycircled{1} captures the tiling. \mycircled{2}, \mycircled{3} and \mycircled{4} show the execution of dense 3$\times$3, 1$\times$1, and depthwise 3$\times$3 convolutions on \neureka{}, respectively.}}
    
    \label{fig:neureka_exec_tiling}
\end{figure*}
\neureka{}'s execution starts when a \riscv{} core offloads a task via configuring a dedicated memory-mapped register. The register file is designed to support queueing two tasks, eliminating extraneous latency due to configuration overhead when running multiple tasks back-to-back.
\neureka{}'s execution is handled via a \gls{fsm}, which controls the dispatch of tile computations on \glspl{pe}. 
\begin{figure*}
    \centering
    \includegraphics[width=0.97\textwidth]{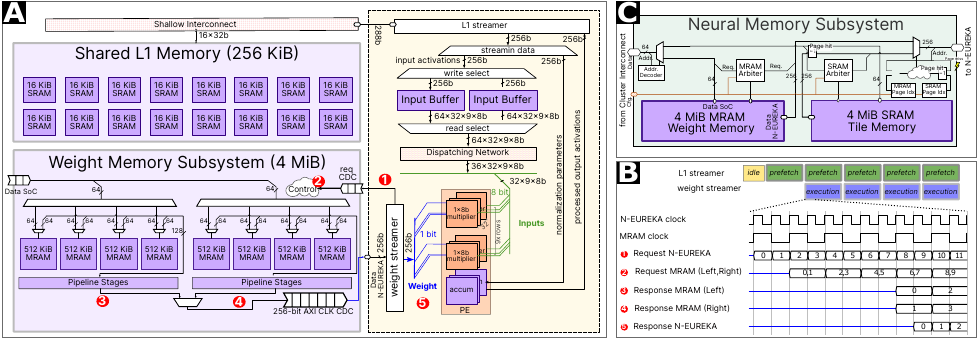}
    \caption{\textit{A)} Detail of the integration of \neureka{} with the MRAM Weight Memory Subsystem.
    \textit{B)} Example of \neureka{} execution, overlapping prefetching through L1 streamer and weight streaming through weight streamer, and detail of the weight streamer operation: \mycircled{1} two weight requests from \neureka{} are propagated through the \gls{cdc} in two cycles and \mycircled{2} propagated to the \gls{mram}; \mycircled{3},\mycircled{4}  the \gls{mram} responds to two requests on parallel banks with latency = 3 internal cycles; \mycircled{5} responses are propagated back to \neureka{} with a total of 9 cycles of latency.
    \textit{C)} Overall architecture of the Neural Memory System with detail of paging mechanism.}
    %
    \label{fig:neureka_exec}
\end{figure*}

The execution flow of \neureka{} follows an output stationary and quasi-input stationary pattern, similar to \cite{conti_marsellus_2023}. The activation layout used by \neureka{} is \gls{hwc} in L1 memory, and weights are packed into 256-bit blocks.
The execution starts with the \textit{prefetch} state by loading 32$\times$8-bit activation per cycle from L1 memory and storing it in the \textsc{InputBuffer}. At the end of the \textit{prefetch} state, the activations are assigned to the multiplier units of all \glspl{pe}  using the \textsc{DispatchingNetwork}. Next, the \textit{execution} state begins with a \SI{256}{\bit\per cycle} access to the \textit{Neural Memory Subsystem} via the weight streamer. The weights fetched are broadcast to all 6$\times$6 \glspl{pe}, allowing the reuse of weights in the spatial dimension. Using the Neural Memory Subsystem to store the \gls{dnn} weights frees the bandwidth towards the L1 memory in the \textit{execution} state. This bandwidth is exploited to prefetch the next activation tile to the second input buffer. This overlap of the \textit{prefetch} and \textit{execution} state allows the accelerator to hide the latency of fetching subsequent tiles.

Figure~\ref{fig:neureka_exec_tiling} illustrates the operation of \gls{dnn} layers on N-EUREKA. The tasks are divided into various tiles executed sequentially. In the dense 3$\times$3 mode, weights corresponding to various bit precisions are fetched in a bit-serial fashion and then summed up within a 32$\times$32-bit accumulator. On the other hand, in the pointwise mode, weights from different precisions are retrieved simultaneously and combined into one of the 32$\times$32-bit accumulators. In the Depthwise mode, the weights are fetched in a bit-serial fashion. However, the accumulators are updated in parallel. In the 3$\times$3 modes, input channel parallelism is exploited in chunks of 28 channels to fit within the available bandwidth (252b of weight per cycle over a limit of 256b). Once all input tiles in the channel dimension for one spatial tile of a convolution have been processed, the partial sum in the accumulators undergoes the \textit{normalization} and \textit{quantization} phase. Then, in the \textit{streamout} state, the normalized and quantized output stored in the accumulators is written back to the L1 memory via the L1 streamer. This process is repeated for all remaining spatial input tiles.

\subsubsection{Weight Memory Subsystem Integration}\label{sec:archi_neureka_wmem}

While conventional memory technologies like \gls{sram} are designed to support high-frequency memory accesses without practical limits on their memory cell lifetime, the write endurance of Spin Transfer Torque (STT) \gls{mram} is limited~\cite{ikegawa_magnetoresistive_2020, guo_spintronics_2021}. Moreover, its write process is orders of magnitude slower than \gls{sram}, taking tens of nanoseconds \cite{ikegawa_magnetoresistive_2020}. Additionally, read access times for STT-\gls{mram} are typically in the range of \SIrange{5}{6}{\nano\second} \cite{lee_331_2023} which corresponds to an operating frequency of \SIrange{160}{200}{\mega\hertz}, which is much lower than the operating frequency of state-of-the-art accelerators \cite{conti_marsellus_2023, houshmand_diana_2023}, making direct integration with high-throughput computing systems like \neureka{} challenging. The proprietary \gls{mram} macro integrated into Siracusa is designed for read access operation at up to \SI{180}{\mega\hertz}.

In Siracusa, we address \gls{mram}'s slow write performance and limited write endurance by integrating it as a static \gls{dnn} weight memory coupled to the accelerator rather than a conventional read-write cache. 
By dimensioning the weight memory subsystem with enough capacity to hold all weights of state-of-the-art \gls{xr} \glspl{dnn} and exploiting its non-volatility, we avoid write accesses to the \gls{mram} at runtime.
Similarly, we address the \gls{mram} IP's slower-than-SRAM read performance by pipelining two word-interleaved memory banks at half the accelerator's clock frequency, as shown in Figure~\ref{fig:neureka_exec}; the banks are accessed in parallel during an \gls{mram} clock cycle (for a total bandwidth of 512 bits/cycle @ 180 MHz). The \gls{dnn} weights are laid out so that the accesses happen in long streams of adjacent addresses, making it possible to hide the access latency of the \gls{mram} cuts due to the analog access time and three internal pipeline stages, as shown in Fig.~\ref{fig:neureka_exec}, on the left.

Following the At-Memory paradigm, weights are directly streamed from the \gls{mram} into \neureka{}'s \glspl{pe}, with minimal intermediate storage in first-in, first-out (FIFO) buffers, matching \neureka{}'s compute throughput.
The \gls{mram} weight memory subsystem can provide \SI{256}{\bit\per cycle}, which is enough to feed the entire input channel parallelism available in \neureka{} (32 input channels) in the 1$\times$1 dense convolution mode, but not in the 3$\times$3 modes.
To align with the available weight memory bandwidth, \neureka{} supports a maximum of 28 input channels in parallel in these modes.

\subsection{Tile Memory vs MRAM memory}
In the memory hierarchy of Siracusa, \gls{mram} and \gls{sram} play distinct roles, each characterized by their unique impacts on speed, power consumption, and area efficiency. \gls{sram} is characterized by its high speed, achieving operational frequencies of 360 MHz and beyond, compared to \gls{mram}'s peak at 180 MHz. Furthermore, \gls{sram} can operate at lower voltages, down to \SI{0.6}{\volt}, providing an edge over \gls{mram}'s \SI{0.65}{\V}. 
On the other hand, \gls{mram} is distinguished by its high density, which is 1.8 $\times$ that of Tile Memory, and its non-volatile nature. These characteristics make \gls{mram} particularly well-suited for applications where space is limited, and data persistence is required, such as in storing weights in Siracusa. The high density of \gls{mram}, at \SI{1.47}{Mi\byte\per\milli\meter\squared}, allows for efficient utilization of space, contributing to the system's overall compactness and efficiency.
To mitigate \gls{mram}'s slower speed, a double buffering technique is employed, effectively boosting bandwidth as explained in Sec.-\ref{sec:archi_neureka_wmem}. With these architectural modifications, Siracusa offers a Peak Area Efficiency of \SI{65.2}{\giga OP\per\second\per\milli\meter\squared}. The \gls{sram} tile memory in Siracusa is predominantly utilized for storing activations, leveraging its superior read and write speeds. This rapid data access capability makes \gls{sram} an ideal choice for handling the dynamic and transient data flows associated with activation values in the \gls{dnn} workloads.
\section{SoC Measurements}\label{sec:meas}

\begin{figure}
\begin{center}
\includegraphics[width=0.95\linewidth]{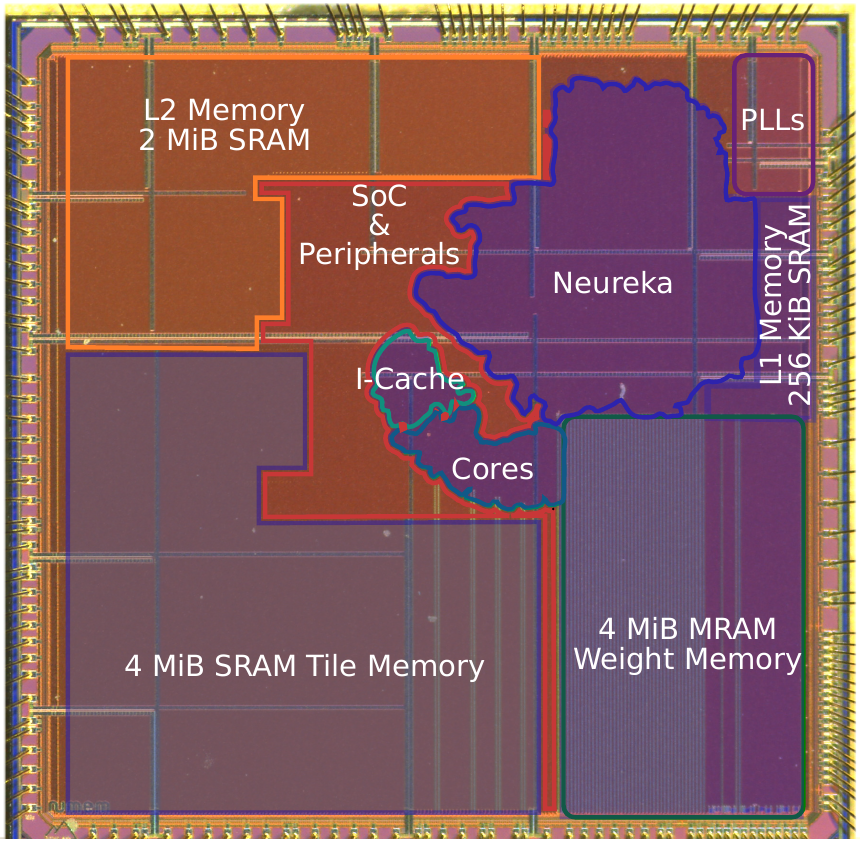}
\caption{Annotated micrograph of a \SI{4}{\milli\meter}$\times$\SI{4}{\milli\meter} Siracusa die. The highlighted \cluster{} components include the \riscv{} cores, L1 memory, instruction cache, \neureka{} and \gls{mram} weight and \gls{sram} tile memories, occupying a total of \SI{10.7}{\milli\meter\squared}. Besides the \cluster{} IPs, the \gls{soc} components, including peripheral controllers, the PLLs and L2 memory, occupying \SI{4.3}{\milli\meter\squared} are highlighted.}
\label{fig:siracusa_die_picmicrograph}
\end{center}
\end{figure}

\begin{figure}
\begin{center}
\includegraphics[width=0.95\linewidth]{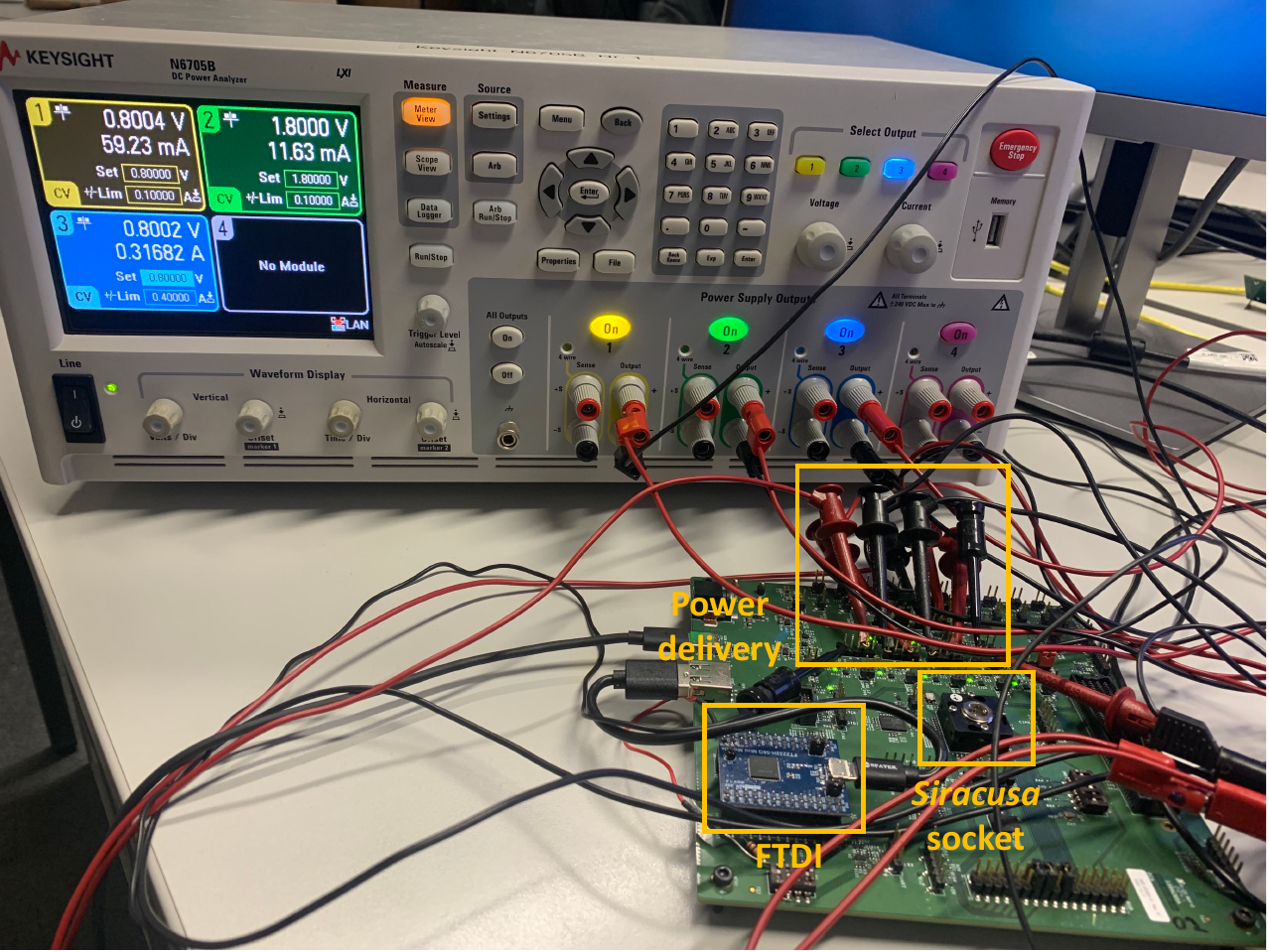}
\caption{Measurement setup for the Siracusa \gls{soc}. The \gls{soc} is embedded in a carrier PCB, which hosts individual pin headers for all power supplies. We used a multi-channel source meter to simultaneously control the power domain's voltages and measure current.}
\label{fig:siracusa_pcb_pic}
\end{center}
\end{figure}

The Siracusa prototype was manufactured in TSMC \SI{16}{\nano\meter} FinFET technology with a total die area of \SI{16}{mm^2} with the \cluster{}'s area occupying \SI{10.7}{\milli\meter\squared}. A micrograph of Siracusa is shown in Figure~\ref{fig:siracusa_die_picmicrograph}.
The fabricated \gls{soc} uses a BGA package mounted in an embedded wire
elastomer socket on a custom evaluation board (Fig.~\ref{fig:siracusa_pcb_pic}), which exposes separate headers for each power domain and allows fine-grained current measurements.
The JTAG interface of the chip, accessible via USB through an FTDI daughterboard, is used to program and debug the \gls{soc}.
This Section introduces the measurement methodology and results measured on the manufactured chip.
Unless otherwise mentioned, we use nominal operating conditions at a core digital supply voltage of \SI{0.8}{\volt} and room temperature; the \gls{mram} array is on a separate voltage domain at \SI{1.8}{\volt}.

\subsection{Frequency/power sweep}\label{sec:power}


\begin{table}
\caption[{Frequency sweep of the \cluster{} and \gls{mram}}]{Frequency sweep of the \cluster{} and \gls{mram}}
\begin{tabularx}{\linewidth}{Xrrrr}
Voltage [V]                   & 0.65 & 0.70 & 0.75 & 0.80 \\
\hline
Cluster Frequency [MHz]         & 210  & 250  & 310  & 360  \\
\hline
Cluster Power (incl. MRAM) [mW] & 151  & 196  & 261  & 332  \\
\hline
MRAM Frequency [MHz]            & 105  & 125  & 155  & 180  \\
\hline
MRAM Power [mW]                & 40   & 47   & 58   & 69  \\

\end{tabularx}
\label{tab:sweep}
\end{table}

\modified{The \gls{mram} IP used has two voltage domains, an analog domain for the array and a digital domain for the memory periphery and interface, which is tied to the \cluster{}'s operating voltage. We determined Siracusa's \cluster{} operating voltage by sweeping the voltage from \SI{0.8}{\V} down to \SI{0.6}{\V}, keeping the \gls{mram} voltage at \SI{1.8}{\V}. During the analysis, we observed that the \gls{mram} could operate at a minimum of \SI{0.65}{\V}. Consequently,} we characterize Siracusa's \cluster{} domain by sweeping the operating voltage between \SI{0.65}{\volt} and \SI{0.8}{\volt} for the digital components (\neureka{}, \riscv{} cores, \gls{mram} peripheral circuitry) without modulating the analog supply voltage sources.
To exercise the design's critical path and determine the maximum stable operating frequency, we use a dense 3$\times$3 Convolution workload with 252 input channels and 32 output channels with a feature map of 6$\times$6.
We chose this workload because it maximally utilizes \neureka{}'s datapath with the available \gls{mram} bandwidth (Section~\ref{sec:archi_neureka_wmem}), while also stressing the bandwidth usage between \gls{mram} and \neureka{}, using an average 85\% of the \gls{mram}$\to$\neureka{} theoretical bandwidth.
Table~\ref{tab:sweep} reports the peak frequency and power as a function of operating voltage measured in the fabricated Siracusa chip. 

The maximum operating frequency measured in the \cluster{} reaches \SI{360}{\mega\hertz} at \SI{0.8}{\volt} where the \gls{mram} operates synchronously at \SI{180}{\mega\hertz}. When the voltage is scaled to \SI{0.65}{\volt}, the maximum operating frequency decreases to \SI{210}{\mega\hertz}. Under nominal conditions, the power consumption of the \cluster{} reaches 
\SI{330}{\milli\watt}, with the \gls{mram} macro's power contributing for 25\,\%. Reducing the \cluster{}'s operating voltage to \SI{0.65}{\volt} and operating frequency to \SI{210}{\mega\hertz} reduces total power consumption by a factor of 2.2$\times$.

\subsection{Performance \& Energy Efficiency}
\subsubsection{RISC-V Cluster Performance}
\begin{figure}
  \begin{center}
\includegraphics[width=0.95\linewidth]{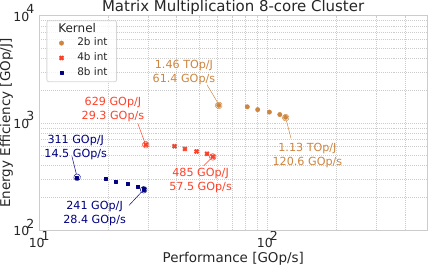}
\caption{Peak energy efficiency versus throughput along the \cluster~ voltage/frequency Pareto frontier, executing matrix multiplication in different precision. The data was measured at room temperature and operating voltages ranging between \SIrange{0.65}{0.8}{\volt} and maximal frequency. \modified{1 MAC = 2Ops is used unless specified otherwise.}}
\label{fig:rv_energy_efficiency}
\end{center}
\end{figure}

We evaluate the throughput and energy efficiency of the octa-core \cluster{}, using optimized matrix multiplication kernels with support for \gls{simd} processing of 2, 4, and 8-bit integer matrices, sweeping the operating voltage from \SIrange{0.65}{0.8}{\volt}. Similarly to the methodology discussed in Section~\ref{sec:power}, we determined the maximum operating frequency of the octa-core compute cluster. We found the maximum operating frequency under nominal conditions to be \SI{530}{\mega\hertz}, and \SI{310}{\mega\hertz} when applying a core voltage of \SI{0.65}{\volt}. Our measurements are plotted in Figure~\ref{fig:rv_energy_efficiency}.

Using the advanced MAC\&LOAD \gls{isa} extensions under nominal conditions, throughput and energy efficiency measurements are \SI{120.6}{\giga Op\per\second}\,@\,\SI{1.13}{\tera Op\per\joule}, \SI{57.5}{\giga Op\per\second}\,@\,\SI{485}{\giga Op\per\joule}, and \SI{28.4}{\giga Op\per\second}\,@\,\SI{241}{GOp\per\joule} for 2, 4, and 8-bit operands. For the energy-efficient \SI{0.65}{\volt} operating corner, energy efficiency increases by 1.3$\times$ while throughput decreases by 2$\times$ for all operand precisions.
\begin{table}[ht]
\centering
\caption{DSP kernels on RISC-V cluster cores @360 MHz}
\label{tab:kernel_comparison}
\adjustbox{max width=\linewidth}{
\begin{tabular}{cccc}
\toprule
Kernel & Data Type & Configuration & Throughput[GFlop/s]  \\ 
\midrule
\multirow{2}{*}{MatMul}  
& FP32  & \multirow{2}{*}{Rows = 64, Columns = 64}  & 1.08    \\ 
& FP16  & & 2.12     \\                     
\midrule
\multirow{2}{*}{KMeans}  
& FP32  & Input Size = 256 & 1.05   \\ 
& FP16  & Features = 8, Clusters = 8 & 1.68     \\                     
\midrule
\multirow{2}{*}{SVM}  
& FP32  & Input Size = 256 & 0.37   \\ 
& FP16  & Kernel Type = Linear & 0.41    \\                     
\midrule
\multirow{2}{*}{FIR}  
& FP32  & Order = 8 & 0.8   \\ 
& FP16  & Length = 4096 & 1.43  \\                     
\midrule
\multirow{2}{*}{FFT}  
& FP32  & \multirow{2}{*}{Input Size = 4096} & 0.21 \\ 
& FP16  & & 0.33  \\  
\midrule
Distortion  & INT  & Image Size = 128 $\times$ 128 $\times$ 3 & 0.26$^{*}$  \\ 
\bottomrule
\end{tabular}
}
\footnotesize\textsuperscript{*} Measured in Gpixels/s , RGB pixel is considered
\end{table}

We also assess the performance of Siracusa across a wide set of \gls{dsp} kernels as captured in Table~\ref{tab:kernel_comparison}. These kernels are derived from the work presented in \cite{mirsalari_translib_2023}, and their applicability is further elaborated in \cite{montagna_low-power_2022}. On average, the kernels attain a throughput of \SI{0.7}{\giga Flops\per\second} when operating with FP32 precision. Using FP16 over FP32 precision enhances the throughput by a factor of 1.7$\times$ (up to \SI{1.2}{\giga Flops\per\second}). While matmul kernel could be used for matrix transformations, we also benchmarked Siracusa on distortion kernel, which is widely used in the post-processing pipeline of \gls{xr}\cite{huzaifa_illixr_2022}. This kernel could achieve a throughput of \SI{0.26}{\giga Pixels\per\second}. Siracusa's capability to accommodate and accelerate a diverse range of applications within the \gls{dsp} pipeline underscores its adaptability for \gls{dsp}-centric tasks pivotal to \gls{xr} applications.

\subsubsection{\neureka{} Performance}

We evaluate the peak throughput and energy efficiency of \neureka{} for \gls{dnn} workloads by measuring representative kernels for each of \neureka{}'s operating modes. Furthermore, we quantify the benefits of the dedicated weight memory subsystem introduced in Section~\ref{sec:archi_neureka_wmem}.
All measurements presented in this Section use statically allocated data in L1 memory and the \gls{mram} memory. These measurements do not consider tiling overheads caused by data movement, typical in large end-to-end networks. An in-depth case study of end-to-end network execution, including system-level overheads caused by tiling, is presented in Section~\ref{results_casestudy}. 

\begin{figure}
  \begin{center}
  \vspace{-0.5cm}
\includegraphics[width=\linewidth]{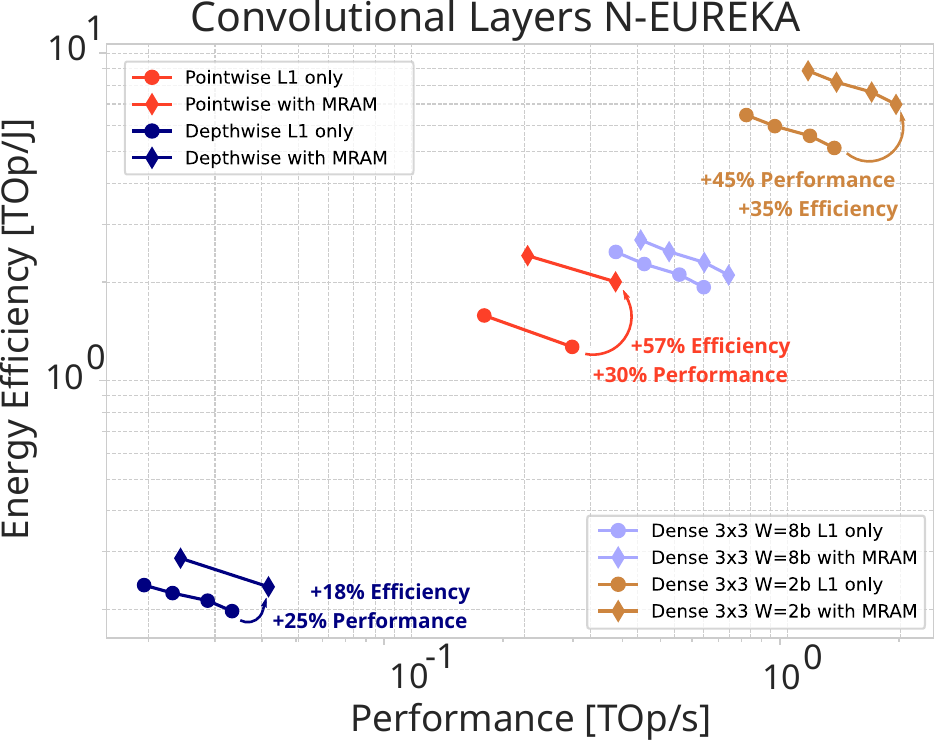}
\caption{{Peak energy efficiency versus throughput of \neureka{} executing depthwise, dense 3$\times$3, and pointwise convolutions using activations from L1 Memory and weights from \gls{mram}. The measurements are taken at room temperature with core voltages between \SI{0.65}{\volt} and \SI{0.8}{\volt}.}}
\label{fig:neureka_perf}
\end{center}
\vspace{-0.5cm}
\end{figure}

We evaluate \neureka{}'s throughput and energy efficiency for all supported operators, i.e., 3$\times$3 dense and depthwise, and 1$\times$1 dense convolutions by sweeping the supply voltage from \SI{0.65}{\volt} to \SI{0.8}{\volt}. We measure layer configurations that optimally utilize the accelerator's datapath; specifically, we use activations with spatial dimensions 6$\times$6, and 224 input channels for the pointwise and depthwise kernels and 252 input channels for the dense 3$\times$3 convolution kernel. Both dense convolutions produce 32 output channels, while the depthwise convolution has 224 output channels.

To quantify the impact of the dedicated weight memory subsystem, we measure two different execution setups; first, we measure the performance using weights allocated in \gls{mram}. As a baseline for comparison, we also measure the performance when fetching weights and activations from L1 memory only. These results are collected in the \hp{} (\SI{0.8}{\V}, \SI{360}{\mega\hertz}) as well as \lp~(\SI{0.65}{\V}, \SI{210}{\mega\hertz}) operating points as shown in Figure \ref{fig:neureka_perf}. 
The results show that for the Depthwise convolution kernel, using \gls{mram} increases both speed and energy efficiency by about 1.2$\times$ compared to just using L1 memory at \hp{} operating point. When operating at a \lp{} operating point, the speed decreases by 1.7$\times$, while the energy efficiency gets slightly better by 1.2$\times$. 
For pointwise, 1$\times$1 kernel, using \gls{mram} leads to a 1.3$\times$ increase in speed and a 1.5$\times$ increase in energy efficiency over L1 memory. \gls{mram} reaches the best energy efficiency of \SI{2.4}{\tera Op \per\joule} at the \lp{} operating point for pointwise kernel. Operating at \hp{} condition,  results in a 1.7$\times$ increase in the throughput with a 1.2 $\times$ decrease in the energy efficiency. 
For dense 3$\times$3 tasks with 8-bit precision, changing to 2-bit precision with only L1 memory more than doubles both speed and energy efficiency, which \gls{mram} further improves by about 1.45 $\times$ for speed and 1.35$\times$ for energy efficiency due to execution of \neureka{} in bit-serial fashion.

The overall peak energy efficiency is attained at \SI{8.84}{TOp\per\joule} at the \lp{} operating point, whereas peak performance is measured under \hp{} conditions, achieving \SI{1947}{GOp\per\second} for 3$\times$3 dense convolution kernel with weight at 2-bit precision. At an 8-bit weight precision, it achieves a peak energy efficiency of \SI{2.68}{TOp\per\joule} at the \lp{} operating condition for the same 3$\times$3 dense convolution kernel. Furthermore, the peak throughput for this kernel is \SI{698}{GOp\per\second} at \hp{} conditions, closely approaching the ideal throughput of \SI{738}{GOp\per\second}.

\section{At-Memory Efficiency: MRAM Integration}\label{results_casestudy}\label{sec:casestudy}


\begin{figure*}
  \begin{center}
\includegraphics[width=\linewidth]{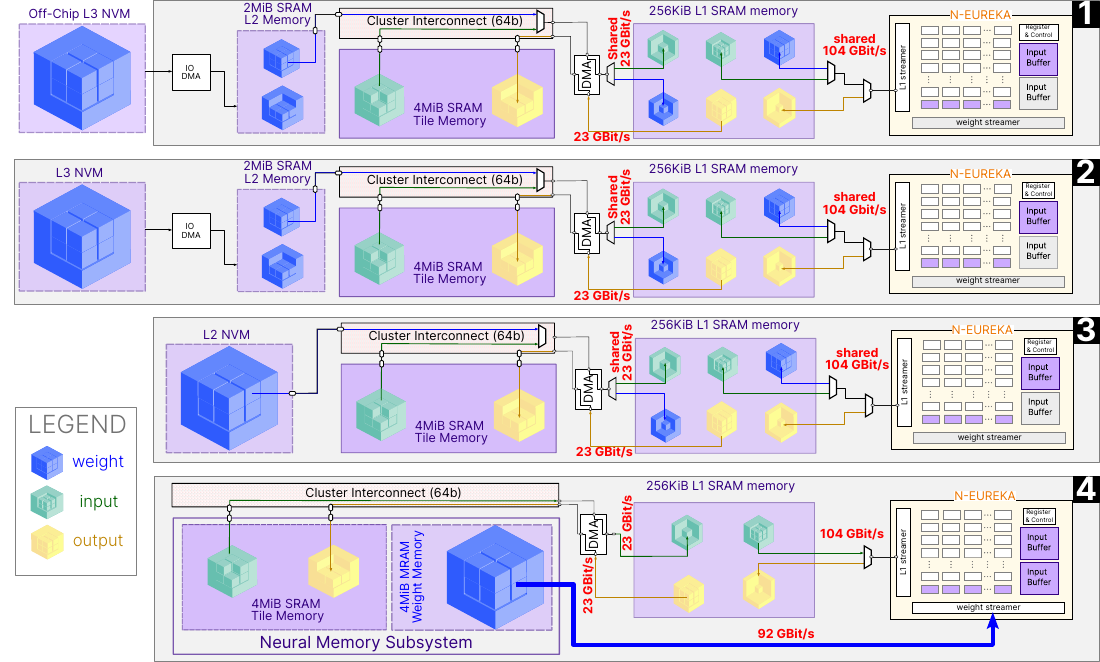}
\caption{Overview of the \gls{nvm} integration scenarios \mycircled{1} \textsc{L3Flash}, \mycircled{2} \textsc{L3MRAM}, \mycircled{3} \textsc{L2MRAM}, and \mycircled{4} \textsc{L1MRAM}. The scenarios are listed in order of progressive tightness of coupling between the \gls{nvm} and the accelerator. In cases \mycircled{1} and \mycircled{2}, \gls{nvm} is used as background L3 memory. In case \mycircled{3}, \gls{mram} is used as L2 memory with equal accessibility for the cores and the accelerator. Finally, in case \mycircled{4} which corresponds to Siracusa's Neural Memory Subsystem, the \gls{mram} is tightly coupled with \neureka{}, providing extra bandwidth for \gls{dnn} weights.}
\label{fig:nvm_integrationcases}
\end{center}
\end{figure*}



\subsection{MRAM integration scenarios}
\label{sec:casestudy_scenarios}
\gls{mram} offers several advantages for \gls{dnn} weight storage over conventional memory technologies like \gls{sram}, as its non-volatility and high memory density allow storing all weights of state-of-the-art \gls{xr} \glspl{dnn} on-chip without copy or transfer overheads.
However, integration into a \gls{soc} is a challenging task, which must address the limited write endurance and read access times slower than SRAM~\cite{ikegawa_magnetoresistive_2020}, which characterize \gls{mram}.
Although several techniques for \gls{nvm}-based in-memory computing have been explored \cite{angizi_mrima_2020, kang_-memory_2017}, to the best of our knowledge, Siracusa is the first edge \gls{soc} employing a tight ``At-MRAM computing'' integration scheme between an all-digital \gls{dnn} accelerator and a \gls{nvm}.
In this Section, we quantitatively justify this design choice, comparing different approaches to integrating \gls{nvm} as on-chip \gls{dnn} weight memory.
Ultimately, we show that close integration with the accelerator allows us to take advantage of the benefits of \gls{mram}, while overcoming its drawbacks.

We evaluated four possible integration strategies for \gls{nvm} captured in Figure~\ref{fig:nvm_integrationcases}, characterized by progressively tighter integration between \gls{nvm} and the execution engine.
All schemes are modeled on the same architecture of Siracusa, including the RISC-V \cluster{} and \neureka{}, except for the integration of the \gls{mram} \gls{nvm}.
The first and baseline scenario \textsc{L3Flash} labelled \mycircled{1} in Figure~\ref{fig:nvm_integrationcases}.
In this scenario, \gls{nvm} is only used for off-chip data storage and is connected to the on-chip memory hierarchy using the \gls{soc} IO-DMA; this case corresponds to systems that exclusively use off-chip Flash memory as background memory.
The second scenario, \textsc{L3\gls{mram}}, labeled \mycircled{2} in Figure~\ref{fig:nvm_integrationcases}, is very similar to \textsc{L3Flash}: instead of relying on off-chip Flash, it employs an on-chip \gls{mram} cut, which enables integration in a FinFet technology node, but with a similar integration scheme.
This is the approach followed by the 22nm Vega~\cite{rossi_vega_2022} \gls{soc} in the integration of \SI{4}{\mebi\byte} of the on-chip \gls{mram}.
The third case, \textsc{L2\gls{mram}}, integrates \gls{mram} with the \gls{soc} domain, allowing equal access to the \gls{fc}, as well as the \cluster{} cores and the accelerator in the \cluster{} domain. 
An approach similar to this is employed by the work of Zhang~et~al.~\cite{zhang_22nm_2022};
their \gls{soc}, which targets visually assisted robot navigation, integrates \SI{2}{\mebi\byte} of \gls{mram} on an AHB-lite interconnect together with a Cortex-M33 core and a Neural Visual Processing Unit.
The final case, \textsc{L1\gls{mram}}, is labelled \mycircled{4} in Figure~\ref{fig:nvm_integrationcases}.
It corresponds to the tightly-coupled Siracusa design in Section~\ref{sec:arch}.
\textsc{L1\gls{mram}} integrates the \gls{mram} closely with the accelerator; in this scenario, the accelerator is given high bandwidth and preferred access to the \gls{mram}.

\subsection{End-to-end DNN latency/energy analysis}
\label{sec:casestudy_endtoend}

To make meaningful comparisons between the four scenarios described in Section~\ref{sec:casestudy_scenarios}, we target the end-to-end deployment of an 8-bit quantized version of MobileNet-V2 \cite{sandler_mobilenetv2_2018} as a realistic large-scale workload. We select MobileNet-V2 as a benchmark due to its widespread use in  various \gls{dnn} based visual processing applications \cite{nagrath_ssdmnv2_2021, li_confidence-aware_2021, hu_automatic_2020}. For instance, in the context of hand-tracking within \gls{xr} \cite{han_megatrack_2020} utilizes layers that are characteristic of MobileNet-V2 within its \gls{cnn} framework.
We use silicon measurements of the HyperBus memory access energy and on-chip memory transfer energy using the \cluster{}-DMA to extract latency and power numbers for the first three scenarios; for the fourth scenario, we rely on direct measurements on the Siracusa prototype.
We use the same operating conditions for all measurements: a high-performance operating point with a core voltage of \SI{0.8}{\volt} and clock frequency of \SI{360}{\mega\hertz}; and a high-efficiency one with a core voltage of \SI{0.65}{\volt} and clock frequency of \SI{210}{\mega\hertz}.
We measured the silicon prototype at room temperature. 

\begin{figure*}
\begin{center}
    \includegraphics[width=0.8\linewidth]{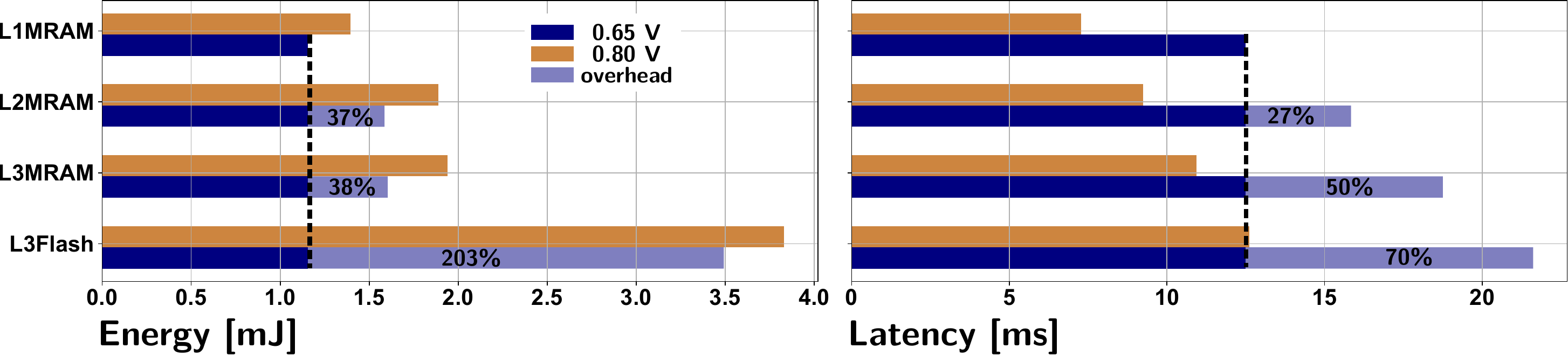}
    \caption{Latency and energy measurements of an inference of MobileNet-V2-1.0-224 on \neureka{} with the four different \gls{nvm} integration scenarios introduced in Section~\ref{results_casestudy}; The \textsc{L1MRAM} scenario, which is integrated in Siracusa, is more efficient, both in terms of latency and energy consumption.}
    \label{fig:nvm_integration1}
\end{center}
\end{figure*}

In the \textsc{L3Flash} case, the \gls{dnn} weights are stored in off-chip flash memory as shown in Figure~\ref{fig:nvm_integrationcases}. During the \gls{dnn} execution, the weights are tiled and transferred from the Flash memory to the \text{L2} memory using the IO-DMA. The \gls{dnn} weight and activation tiles are transferred through a shared \cluster{}-DMA to the \text{L1} memory. \neureka{} accesses the weights and the activations through a high bandwidth, \SI{104}{\giga\bit/\second}, interconnect to the \text{L1} memory for execution of \gls{dnn} inference. 
After the execution on a tile has ended, the output activations are moved to the \text{L2} memory using the \cluster{}-DMA.
We use double buffering for both weights and activations between all memory levels to minimize memory transfer latency overheads.
The results of this analysis, which compares latency and energy cost per inference of MobileNet-V2 between all scenarios, are shown in Figure~\ref{fig:nvm_integration1}.

It takes \SI{12.6}{\milli\second} per frame for the architecture in \textsc{L3Flash} to perform a MobileNet-V2 inference with an energy consumption of \SI{3.8}{\milli\joule}.
The off-chip memory access energy comprised of the 55\% of the total execution energy. The effect on energy is even more pronounced since the cost of moving data from off-chip memory is significant compared to on-chip data transfer: reducing the operation voltage to \SI{0.65}{\volt} and decreasing the operating frequency proportionately does not result in significant energy savings, as the main contributing factor is off-chip memory transactions between the ASIC and the Flash memory that do not scale with the lower core voltage.

The architecture for \textsc{L3\gls{mram}}, using \gls{mram} as an on-chip \text{L3} memory to store the \gls{dnn} weights, was proposed in \cite{rossi_vega_2022}. While the tiled execution scheme remains unchanged with respect to the \textsc{L3Flash} scenario, using \gls{mram} as an on-chip \text{L3} memory lowers end-to-end inference energy by a factor of 2$\times$ as shown in Figure~\ref{fig:nvm_integration1}. These energy savings are attributed to significantly lower data movement energy costs from on-chip \gls{mram}$\,\to\,$\text{L2} compared to the off-chip memory accesses in the \textsc{L3Flash} scenario. However, inference latency improves only by 20\,\% compared to \textsc{L3Flash} as the bandwidth bottleneck between \text{L3}$\,\to\,$\text{L2} still affects throughput.

In the \textsc{L2\gls{mram}} scenario, the \gls{mram} supports direct access from the cores and the accelerator.
Concerning neural network inference, this organization eliminates one level of tiling for weights, reducing congestion at the \text{L3}$\,\to\,$\text{L2} HyperBus interface \cite{rossi_vega_2022}.
In this scenario, the weight and activation are tiled from \text{L2} \gls{mram} and \text{L2} \gls{sram} respectively and transferred to the \text{L1} memory using the shared \cluster{}-DMA, supporting a sustained bandwidth of \SI{23}{G\bit\per\second}. This removes the bandwidth bottleneck from the \text{L3}$\,\to\,$\text{L2} \gls{sram} memory found in \textsc{L3Flash} and \textsc{L3MRAM}, improving the end-to-end latency by 1.2$\times$. The energy savings obtained are negligible compared to \textsc{L3\gls{mram}} since the computing energy becomes dominant constituting approximately 85\% of the total energy once off-chip transfers are removed.

In the last case, \textsc{L1MRAM}, \gls{mram} is tightly integrated with \neureka{}, which improves latency and decreases the inference energy cost compared to all other scenarios.
The tightly-coupled \gls{nvm} integration has two main advantages compared to the previous \textsc{L2MRAM} case.
The first one is given by the reduced pressure on the \text{L2}$\,\to\,$\text{L1} transfers performed by the \cluster{}-DMA, whose bandwidth can be fully dedicated to activations as weights employ the dedicated \SI{92}{\giga\bit/\second} connection between weight memory and \neureka{}.
The second advantage is related to the ability to enable the \neureka{} input buffer prefetching mechanism described in Section~\ref{sec:archi_neureka}; execution can be fully overlapped with input prefetching thanks to the separate input/activation memory ports.
Combined, these lead to a 27\,\% improvement in latency and 37\,\% improvement in energy compared to the \textsc{L2MRAM} scenario.
In particular, compared to the \textsc{L3Flash} case, latency and energy are reduced by 1.7$\times$ and 3$\times$.
Overall, MobileNet-V2 can be run within a latency of \SI{7.3}{\milli\second} and an energy budget of \SI{1.4}{\milli\joule}. If we consider a realistic target frame rate of 30FPS, this is enough to bring power consumption down to $<$\SI{60}{\milli\watt}, matching our application target.
The power is obtained by assuming Siracusa at idle mode operating at \SI{0.65}{\V} and \gls{mram} with negligible power consumption. 

\begin{figure}
\begin{center}
    \includegraphics[width=\linewidth]{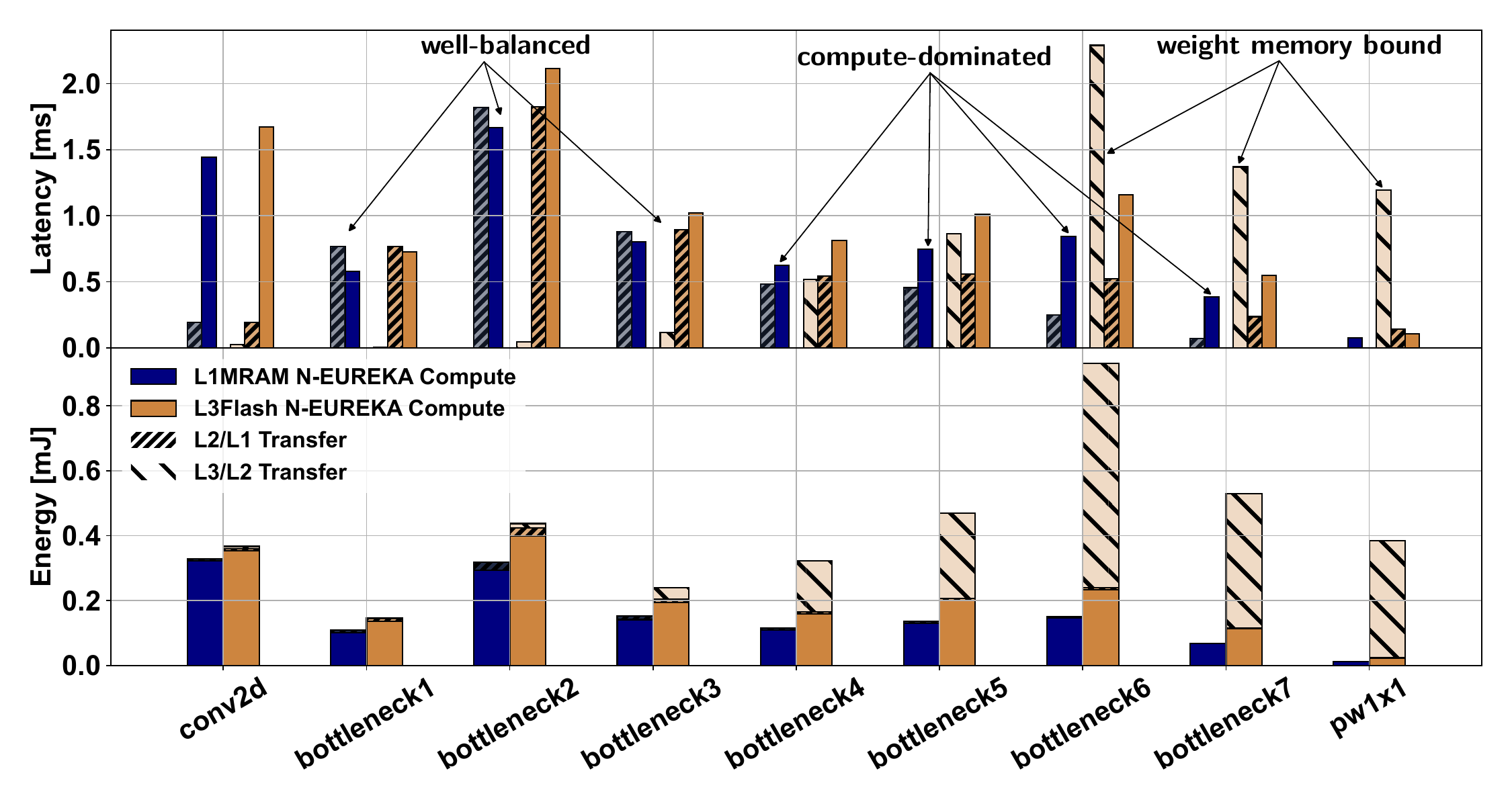}
    \caption{MobileNet-V2-1.0-224 end-to-end execution with \gls{nvm} integration, comparing the \textsc{L3Flash} and \textsc{L1\gls{mram}} scenarios introduced in Section~\ref{results_casestudy}. Each layer's latency measurements are further broken down into compute and memory transfer steps, with the slowest step's latency determining the overall latency. All measurements are obtained at an operating frequency of \SI{360}{\mega\hertz} at \SI{0.8}{\volt} at room temperature.}
    \label{fig:mobilenetv2_perf1}
\end{center}
\end{figure}

\subsection{Layer-wise DNN latency/energy analysis}
\label{sec:casestudy_layerwise}
We further evaluate the impact of \gls{mram} integration by analyzing the effect on each network layer in detail.
We focus our analysis on latency and energy in the \textsc{L1MRAM} and \textsc{L3Flash} scenarios, as shown in Figure~\ref{fig:mobilenetv2_perf1}. We use double-buffering on every memory hierarchy level and report the accumulated latency of memory transfer and compute steps during each layer's execution; as such, the overall latency is determined by the latency of the slowest step.
We find three distinct regimes of operation: in the \textit{well-balanced} regime, found in the first three bottleneck layers, the overall latency is approximately equally distributed between compute and memory access.
In the \textit{compute-dominated} regime, the latency of computations outweighs the latency of memory transfers.
The last regime we distinguish is the \textit{weight memory bound} regime, where latency and energy consumption are dominated by memory transfers for the \textsc{L3Flash} scenario.
Figure~\ref{fig:mobilenetv2_perf1} shows a clear trend; in deeper network layers, where the ratio between spatial dimensions and filter dimensions, and thus weight reuse, is lower, the benefit of a dedicated weight memory subsystem becomes evident.
The energy savings in the sixth bottleneck layer are especially notable. By eliminating $\text{L3}\,\to\,\text{L2}$ transfers for weights, we reduce the energy consumption of the layer by 6.5$\times$.

We can trace the performance improvements unlocked by our architecture to three key design considerations in Siracusa. First, coupling the \gls{mram} directly with \neureka{} allows us to maximize the available bandwidth towards the accelerator. In Siracusa, the weight memory subsystem doubles the available bandwidth of the accelerator in the \cluster{} domain to \SI{184}{\giga\bit/\second}. Second, integrating \gls{mram} sized to hold all the weights of the most common \glspl{dnn} on the chip drastically reduces the energy spent on data movement. While we measure an almost 2$\times$ reduction of energy cost for all \gls{mram}-based designs by eliminating off-chip memory transfers, eliminating all on-chip weight transfers further reduces the end-to-end energy per inference by 1.38$\times$. Third, by keeping the weights statically in the weight memory subsystem and eliminating weight accesses from the \gls{tcdm}, additional L1 bandwidth is available for activation data movement. Although we find that eliminating the \text{L3}$\,\to\,$\text{L2} bottleneck decreases latency by 1.15 - 1.36$\times$, unlocking extra bandwidth towards the last memory level further increases throughput by an additional 27\,\%.

\section{SoA Comparison}\label{sec:discussion}
\label{fig:SoA}

\begin{table*}[b]
\caption{State of the Art Comparison of Edge and near-sensor Computing Systems. Best values are highlighted. }
\label{tab:soacomparison}
  \adjustbox{max width=\linewidth}{
  \begin{tabular}{ccccccc}\toprule
        \phantom{a} & Vega\cite{rossi_vega_2022} & DIANA\cite{houshmand_diana_2023} & MARSELLUS\cite{conti_marsellus_2023}  & Chang et al.\cite{chang_40nm_2022}  & Zhang et al.\cite{zhang_22nm_2022}&\bf{THIS WORK} \\
    \midrule
        \textbf{Technology} & 22nm FDX & 22nm FDX & 22nm FDX & 40nm &  22nm & 16nm FinFET \\
        \multirow{2}{*}{\textbf{Area}} & \multirow{2}{*}{\SI{10}{\milli\meter\squared}} & \multirow{2}{*}{\SI{10.24}{\milli\meter\squared}} & \multirow{2}{*}{\SI{8.7}{\milli\meter\squared}} & \multirow{2}{*}{\SI{25}{\milli\meter\squared}} &  \multirow{2}{*}{\SI{8.76}{\milli\meter\squared}} & \SI{16}{\milli\meter\squared} \\
&&&&&&\cluster{}: \SI{10.7}{\milli\meter\squared}, \iodomain{}: \SI{4.3}{\milli\meter\squared}
        \\
        \textbf{Applications} & IoT GP+DNN SoC & AI-IoT SoC & IoT GP+AI-IoT SoC & Edge AI ASIC & GP+DNN Robot SoC & XR SoC \\
    \midrule
        \multirow{3}{*}{\textbf{Cores}} & 10$\times$ RV32IMCFXpulp & 1$\times$ RV32IMCFXpulp& 1$\times$ RV32IMCFXpulp & 1$\times$ 1$\times$ Cortex-M3 & Cortex-M33 & 1$\times$ RV32IMCFXpulp \\
        & +Convolution Engine & +digital NE& +16$\times$ RV32IMCFXpulpnn & +MAC Array & +NVPU& +8$\times$ RV32IMCFXpulpnn \\
         &  & +AIMC SRAM& +NE &  & & +\neureka \\
    \midrule
       \multirow{2}{*}{\textbf{On-Chip SRAM}} & \SI{128}{\kibi\byte} (L1) & \SI{896}{\kibi\byte} & \SI{128}{\kibi\byte} (L1)  & \multirow{2}{*}{\SI{768}{\kibi\byte}} & \multirow{2}{*}{\SI{1428}{\kibi\byte}} & \SI{256}{\kibi\byte} (L1)\\ 
       & \SI{1600}{\kibi\byte} (L2) & (in/out+weights) & \SI{1024}{\kibi\byte} (L2)  &  &  & +\SI{2}{\mebi\byte} (L2) + \SI{4}{\mebi\byte} (\gls{sram})\\ 
    \midrule
        \textbf{On-chip NVM}  & \SI{4}{\mebi\byte} MRAM (L3) & - & -  & \SI{2.25}{\mebi\byte} RRAM & \SI{2}{\mebi\byte} (L2) & \SI{4}{\mebi\byte} MRAM (L1)\\  
    \midrule
        \multirow{2}{*}{\textbf{INT Precision}} & \multirow{2}{*}{8, 16, 32} & \multirow{2}{*}{2, 4, 8, 16, 32} & 2, 4, 8, 16, 32 (\riscv)  & 1, 4, 8 (MAC Array) & \multirow{2}{*}{16} &  2, 4, 8, 16, 32 (\riscv)\\ 
        &  &  & 2-8 (RBE)  & 32 (Cortex-M3) &  & 2-8b W, 8b in/out (\neureka)\\
    \midrule
        \textbf{Supply Voltage}  & \SIrange{0.5}{0.8}{\V}& \SIrange{0.5}{0.9}{\V} & \SIrange{0.5}{0.8}{\V} & \SI{0.9}{\volt} & \SIrange{0.5}{1.0}{\V} & \SIrange{0.65}{0.8}{\V}\\  
    \midrule
        \textbf{Max Frequency}  & \SI{450}{\mega\hertz}& \SI{320}{\mega\hertz} & \SI{420}{\mega\hertz} & \SI{200}{\mega\hertz} & \SI{56}{\kilo\hertz} to \SI{190}{\mega\hertz} & \SI{360}{\mega\hertz}\\  
    \midrule
         \textbf{Power Range}  & \SI{1.7}{\micro\watt} to \SI{49.4}{\milli\watt}
        & \SIrange{10}{129}{\milli\watt} (digital NE) & \SIrange{12.8}{123}{\milli\watt} & \SIrange{2.6}{131}{\milli\watt} & \SI{468}{\micro\watt} to \SI{158}{\milli\watt} & \SIrange{151}{332}{\milli\watt}\\ 
    \midrule
        \textbf{Peak Performance (8-bit)} & \SI{32.2}{\giga Op\per\second} & \SI{140}{\giga Op\per\second} (digital NE) & \SI{90}{\giga Op\per\second} & N/A & \SI{146}{\giga Op \per\second} (16b in/W) &\done{\SI{698}{\giga Op\per\second}}\\
    \midrule
        \textbf{Peak Efficiency} & \multirow{2}{*}{\SI{1.3}{\tera Op\per\joule}} & \multirow{2}{*}{\SI{2.07}{\tera Op\per\joule}} & \multirow{2}{*}{\SI{1.8}{\tera Op\per\joule}} & \multirow{2}{*}{\SI{0.94}{\tera Op\per\joule}} & \SI{0.7}{\tera Op \per\joule} & \multirow{2}{*}{\done{\SI{2.68}{\tera Op\per\joule}}}\\
         \textbf{(8-bit, no sparsity)} &  & & & &  (16b in/W, \SI{3.5}{\tera Op\per\joule} @ 80\% W sparsity) & \\
    \midrule    
        \textbf{Peak Performance} & \multirow{2}{*}{\SI{32.2}{\giga Op\per\second}} & \SI{140}{\giga Op\per\second} (digital NE) & \SI{637}{\giga Op\per\second} & \multirow{2}{*}{N/A} & \SI{146}{\giga Op \per\second} &\done{\SI{1.95}{\tera Op\per\second}}\\
        \textbf{(Best)} &  & \SI{29.5}{\tera Op\per\second} (AIMC-SRAM) & (2b in, 2b W) & & (16b in/W) & (8b in, 2b W)\\
    \midrule    
        \textbf{Peak Efficiency} & \SI{1.3}{\tera Op\per\joule} & \SI{4.1}{\tera Op\per\joule} (digital NE, 2b in/W) & \SI{12.4}{\tera Op\per\joule} & \done{\SI{60.64}{\tera Op\per\joule}} &  \SI{0.7}{\tera Op \per\joule} & \SI{8.84}{\tera Op\per\joule}\\
        \textbf{(Best, no sparsity)} & (8b in/W) & \done{\SI{600}{\tera Op\per\joule}} (AIMC-SRAM) & (2b in/W) & (1b in/W) & (16b in/W, \SI{3.5}{\tera Op\per\joule} @ 80\% W sparsity) & (8b in, 2b W)\\
    \midrule    
        \textbf{Peak Binary} & \SI{83.2}{\tera Bop\per\joule} & \multirow{2}{*}{\SI{16.4}{\tera Bop\per\joule} (digital NE, 2b in/W)} & \SI{49.6}{\tera Bop\per\joule} & \SI{60.64}{\tera Bop\per\joule} &  \done{\SI{179}{\tera Bop \per\joule}} & \textcolor{black}{{\SI{141.4}{\tera BOp\per\joule}}}\\
        \textbf{Equivalent Efficiency} & (8b in/W) &  & (2b in/W) & (1b in/W) & (16b in/W) & (8b in, 2b W)\\
    \midrule    
        \textbf{Peak Area Efficiency} & \SI{3.2}{\giga Op\per\second\per\milli\meter\squared} & \SI{21.2}{\giga Op\per\second\per\milli\meter\squared} (digital NE) & \SI{47.4}{\giga Op\per\second\per\milli\meter\squared} (cluster) & N/A & \SI{58.3}{\giga Op\per\second\per\milli\meter\squared} & \done{\SI{65.2}{\giga Op\per\second\per\milli\meter\squared}} (cluster)\\
    \midrule   
  \end{tabular}}
\end{table*}

Emerging near-sensor \gls{xr} requires combining high performance in crucial \gls{ai} kernels with flexibility and low power footprint to enable integration in wearable devices.
In Table~\ref{tab:soacomparison}, we compare Siracusa's performance and efficiency with that of state-of-the-art computing platforms for near-sensor \gls{xr}, accelerated extreme-edge \gls{ai}, and near-sensor computing -- using a variety of technologies such as embedded \gls{nvm}~\cite{rossi_vega_2022,zhang_22nm_2022}, \gls{aimc}~\cite{houshmand_diana_2023} and digital acceleration~\cite{conti_marsellus_2023,chang_40nm_2022,zhang_22nm_2022}.
While Zhang~et~al.~\cite{zhang_22nm_2022} use a non-standard factor of 1 16-bit-MAC = 7 Ops in their paper, we report all results considering 1 MAC = 2 Ops at native precision supported by the chip for fairness following the guidelines in \cite{burr_fair_2022}.
Although there is a rich body of literature to motivate aggressive pruning of neural networks, the impact of sparsity on the energy efficiency of accelerators is mainly dependent on the distribution of zeros; however, accounting for the structure of sparsity in \gls{dnn} weights is nearly impossible \cite{hoeer_sparsity_2021}. Therefore, to fairly compare the energy efficiency of different designs, we chose to normalize all reported efficiency numbers to 0\,\% weight sparsity.

To the best of the authors' knowledge, Siracusa is the first SoC to integrate non-volatile \gls{mram} closely coupled to an all-digital \gls{dnn} accelerator.
Siracusa achieves the best peak performance of \SI{1.95}{TOp\per\second} (8b activation, 2b weight configuration) compared to other works \cite{rossi_vega_2022, conti_marsellus_2023, chang_40nm_2022, zhang_22nm_2022} except DIANA\cite{houshmand_diana_2023} 's \gls{sram}-based \gls{aimc} accelerator.
Regarding 8-bit peak performance, it provides the absolute highest performance compared to the other works in Table~\ref{tab:soacomparison}, due to the large embedded \neureka{} engine.

Considering energy efficiency, Siracusa achieves the highest peak efficiency at 8-bit precision compared to all \glspl{soc} considered when normalizing for 0\,\% sparsity. Neglecting this normalization, Zhang~et~al.~\cite{zhang_22nm_2022} report a 31\,\% higher overall energy efficiency than our work; however, these results are obtained with a very high sparsity rate (80\,\% for weights, 50\,\% for activations).
Finally, when considering the best overall efficiency (i.e., at any precision), Siracusa achieves \SI{8.84}{\tera Op/\joule} with 2-bit weights and 8-bit activations.
This compares favourably to other digital \glspl{soc}; even when considering instances where the absolute efficiency reported is better, Siracusa is competitive regarding energy per elementary operation.
We can highlight this fact by considering equivalent binary operations (Bops)\footnote{
Bops = $N_{bits,in}\times N_{bits,W}\times Ops$
}: Siracusa achieves \SI{141}{\tera Bop/\joule};  Marsellus~\cite{conti_marsellus_2023} achieves \SI{49.6}{\tera Bop/\joule}; and Chang~et~al.~\cite{chang_40nm_2022} achieve \SI{60.6}{\tera Bop/\joule}.
The only digital \gls{soc} achieving higher binary efficiency than Siracusa is Zhang~et~al.'s~\cite{zhang_22nm_2022} (\SI{179}{\tera Bop/\joule}): in their case, the choice to use 16-bit inputs and weights boosts this particular metric, due to the more accurate numerical representation.
The \gls{aimc} accelerator in DIANA offers 67$\times$ better peak energy efficiency compared to Siracusa; however, exploiting this design to high utilization and full efficiency is very complex according to DIANA's own architects~\cite{van_delm_htvm_2023}, and this efficiency comes at a steep price in terms of noise and accuracy.
At 8-bit precision, Siracusa achieves the highest peak efficiency compared to all considered \glspl{soc}.

The embedded \gls{mram} \gls{nvm} and its tight integration with the accelerator set \neureka{} apart from all other \glspl{soc} of similar class.
As discussed in Section~\ref{sec:casestudy}, a large \gls{nvm} is necessary to run common edge \glspl{dnn} fully on-chip, and the L1 tight integration strategy employed in Siracusa maximizes performance gains in end-to-end execution.
Of the considered \glspl{soc}, only Vega~\cite{rossi_vega_2022} and Siracusa include enough on-chip \gls{mram} to support the execution of a MobileNet-V2 network~\cite{sandler_mobilenetv2_2018} at 8-bit precision.
We observe that as \glspl{nvm} are typically not symmetric in read and write power and latency, it is not possible to use them as on-chip cache for a larger off-chip weight memory; therefore, the other considered \glspl{soc} need to use off-chip transfers to run common edge \glspl{dnn}, suffering from a heavy penalty on both end-to-end latency and efficiency.

Finally, thanks to the high-density on-chip integration of \gls{mram} near \neureka{}, Siracusa achieves the best area efficiency compared to the existing SoA architectures\cite{rossi_vega_2022, houshmand_diana_2023, conti_marsellus_2023} -- a prerequisite to enabling high-performance and low power in a constrained form factor and overall cost as necessary for \gls{xr} wearable devices. 
\section{Conclusion}\label{sec:conclusion}
In this work, we presented three key ideas to integrate emerging \gls{mram} within a state-of-the-art \gls{xr} near-sensor computing system, and evaluated their impact in terms of energy efficiency and end-to-end latency of realistic \glspl{dnn}. 
The key innovations demonstrated in this work are:
1) Closely coupling non-volatile \gls{mram} as dedicated weight memory with the system's accelerator unlocks additional bandwidth for weight transfers during network inference, doubling the effective bandwidth for layers with low weight reuse.
2) Leveraging the high memory density of \gls{mram} enables all-weights-on-chip inference, drastically reducing the energy cost of data movement for \gls{dnn} inference and consequently improving end-to-end energy efficiency.
3) Careful design of the accelerator's weight memory access pattern and integration of a low-overhead \gls{mram} pipelining system allows us to prefetch network weights in a non-speculative manner, fully compensating for the slower-than-SRAM read access time of \gls{mram} without impacting energy efficiency.

The combined effect of these key innovations improves the end-to-end inference latency of \gls{dnn} workloads by 1.7$\times$ and reduces their energy cost by 3$\times$ compared to conventional systems using off-chip \gls{nvm}. While pure \gls{aimc} macros achieve higher peak efficiency than our all-digital accelerator, we demonstrate end-to-end deployment results of realistic \gls{dnn} workloads, achieving a throughput of \SI{698}{\giga Op\per\second} on 8-bit quantized networks, 4.8$\times$ more than the highest value reported in \gls{xr} \gls{soc} literature, while achieving state-of-the-art energy efficiency of \SI{8.84}{\tera Op\per\joule}. Thanks to the tight integration of high-density \gls{mram} with \neureka{}, Siracusa's \cluster{} improves on the state-of-the-art in compute area efficiency by 10\,\% while integrating \SI{10.25}{\mebi\byte} of memory, 1.8$\times$ more than Vega \cite{rossi_vega_2022}, the design with the largest amount of on-chip memory.

\bibliographystyle{IEEEtran}
\bibliography{references}

\begin{IEEEbiography}[{\includegraphics[width=0.95in,height=1.25in,clip,keepaspectratio]{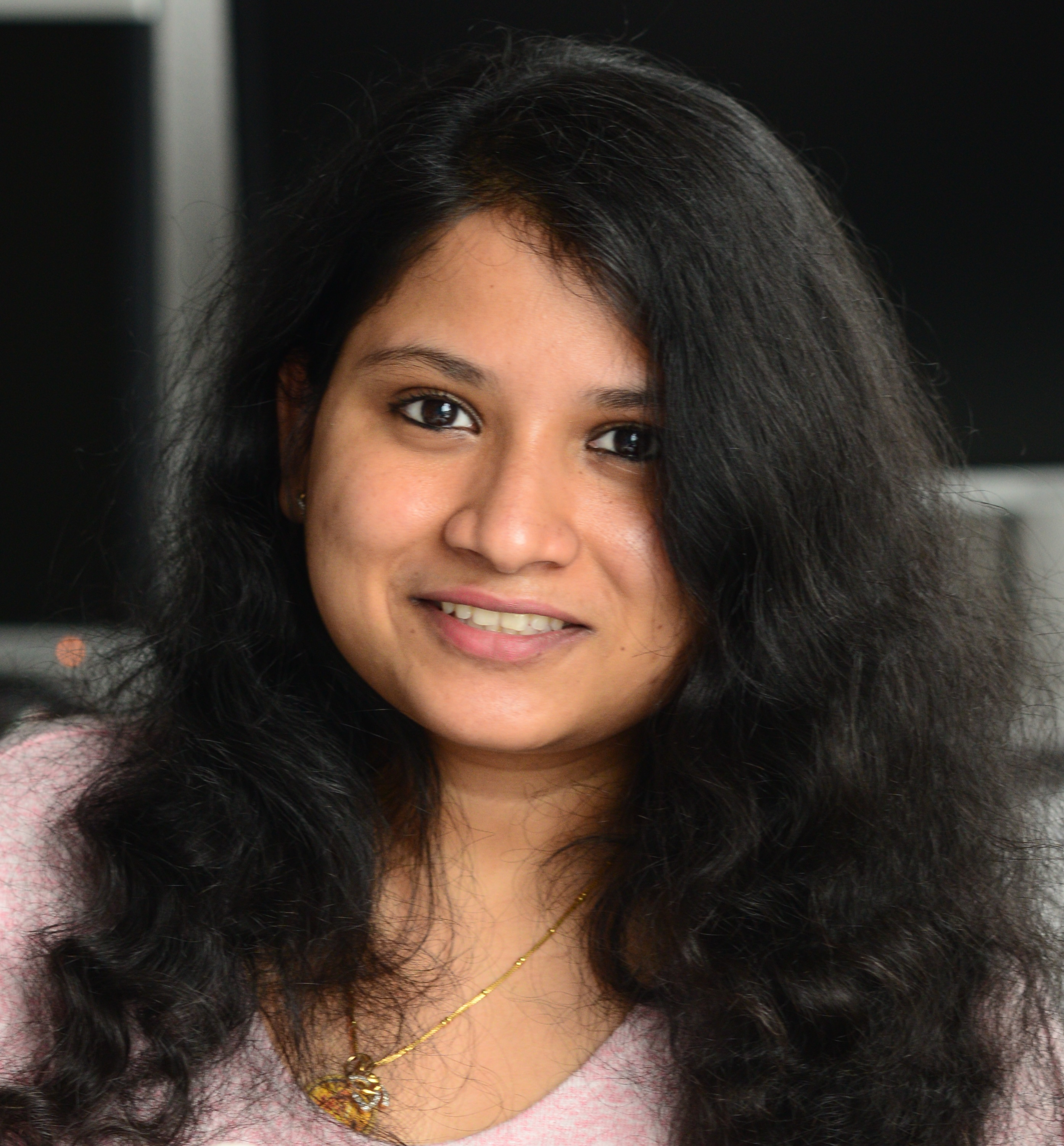}}]{Arpan Suravi Prasad} (Graduate Student Member, IEEE) received B.Tech degree in electronics and communication engineering from the National Institute of Technology(NIT), Rourkela, India in 2015 and received an M.Sc. degree in electrical engineering and information technology from ETH Z\"urich in 2021, where she is currently pursuing a Ph.D. degree with the Integrated Systems Laboratory under the supervision of Prof. Luca Benini. Her research interests include designing energy-efficient hardware architectures targeting extended reality (XR) edge applications and machine learning algorithms.
\end{IEEEbiography}

\begin{IEEEbiography}[{\includegraphics[width=0.95in,height=1.25in,clip,keepaspectratio]{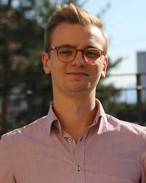}}]{Moritz Scherer} (Graduate Student Member, IEEE)
received the B.Sc. and M.Sc. degree in electrical engineering and information technology from ETH Zürich in 2018 and 2020, respectively, where he is currently pursuing a Ph.D. degree at the Integrated Systems Laboratory. His current research interests include the design of ultra-low power and energy-efficient circuits and accelerators as well as system-level and embedded design for machine learning and edge computing applications.
\end{IEEEbiography}

\begin{IEEEbiography}[{\includegraphics[width=0.95in,height=1.25in,clip,keepaspectratio]{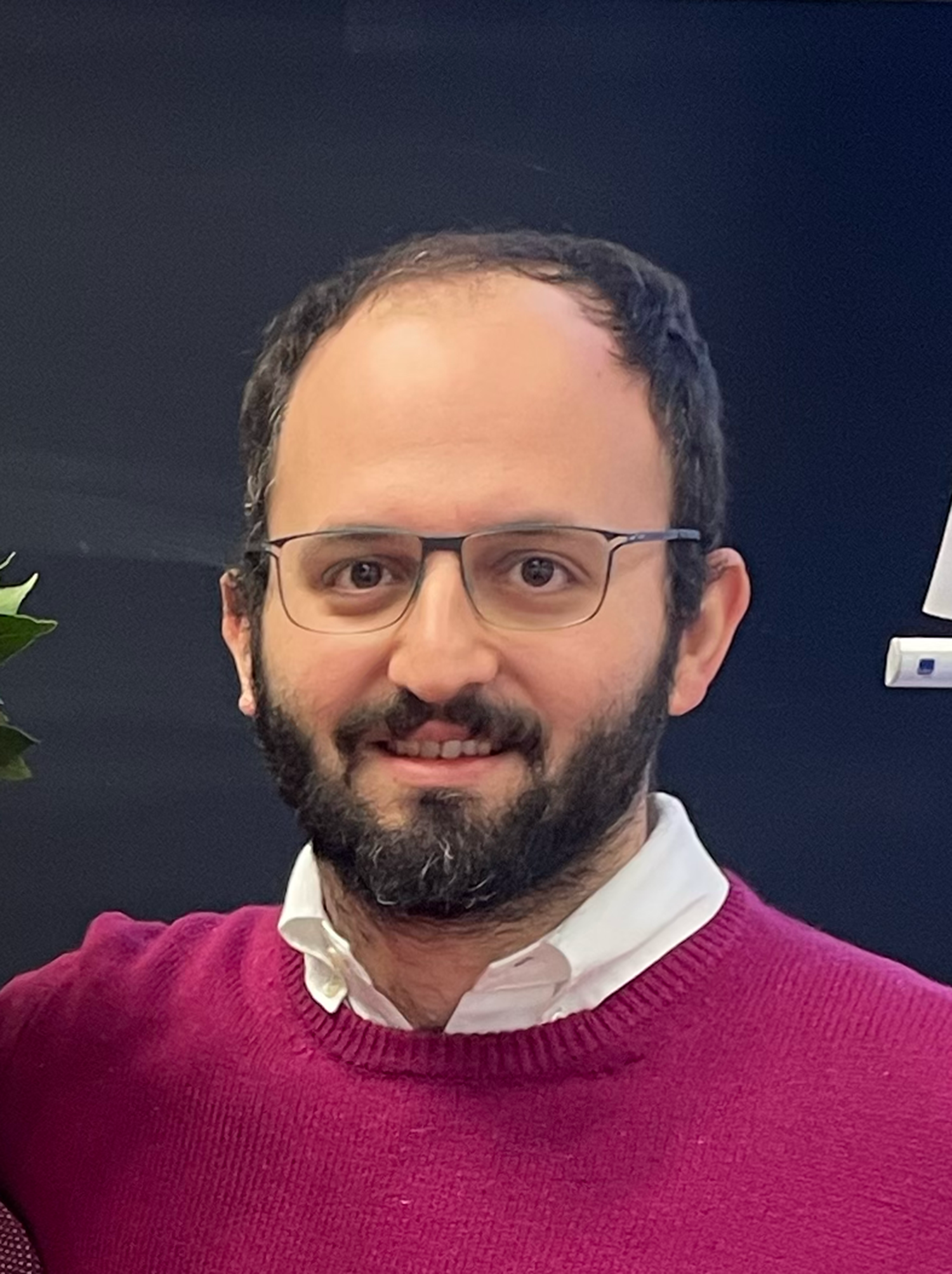}}]{Francesco Conti} (Member, IEEE) received the Ph.D. degree in electronic engineering from the University of Bologna, Italy, in 2016.
He is currently a Tenure-Track Assistant Professor with the DEI Department, University of Bologna. From 2016 to 2020, he held a research grant with the University of Bologna and a position as Post-Doctoral Researcher with ETH Zürich.
His research is centered on hardware acceleration in ultra-low power and highly energy efficient platforms, with a particular focus on System-on-Chips for Artificial Intelligence applications.
His research work has resulted in more than 90 publications in international conferences and journals and was awarded several times, including the 2020 IEEE \textsc{Transactions on Circuits and Systems I: Regular Papers} Darlington Best Paper Award.
\end{IEEEbiography}

\begin{IEEEbiography}[{\includegraphics[width=0.95in,height=1.25in,clip,keepaspectratio]{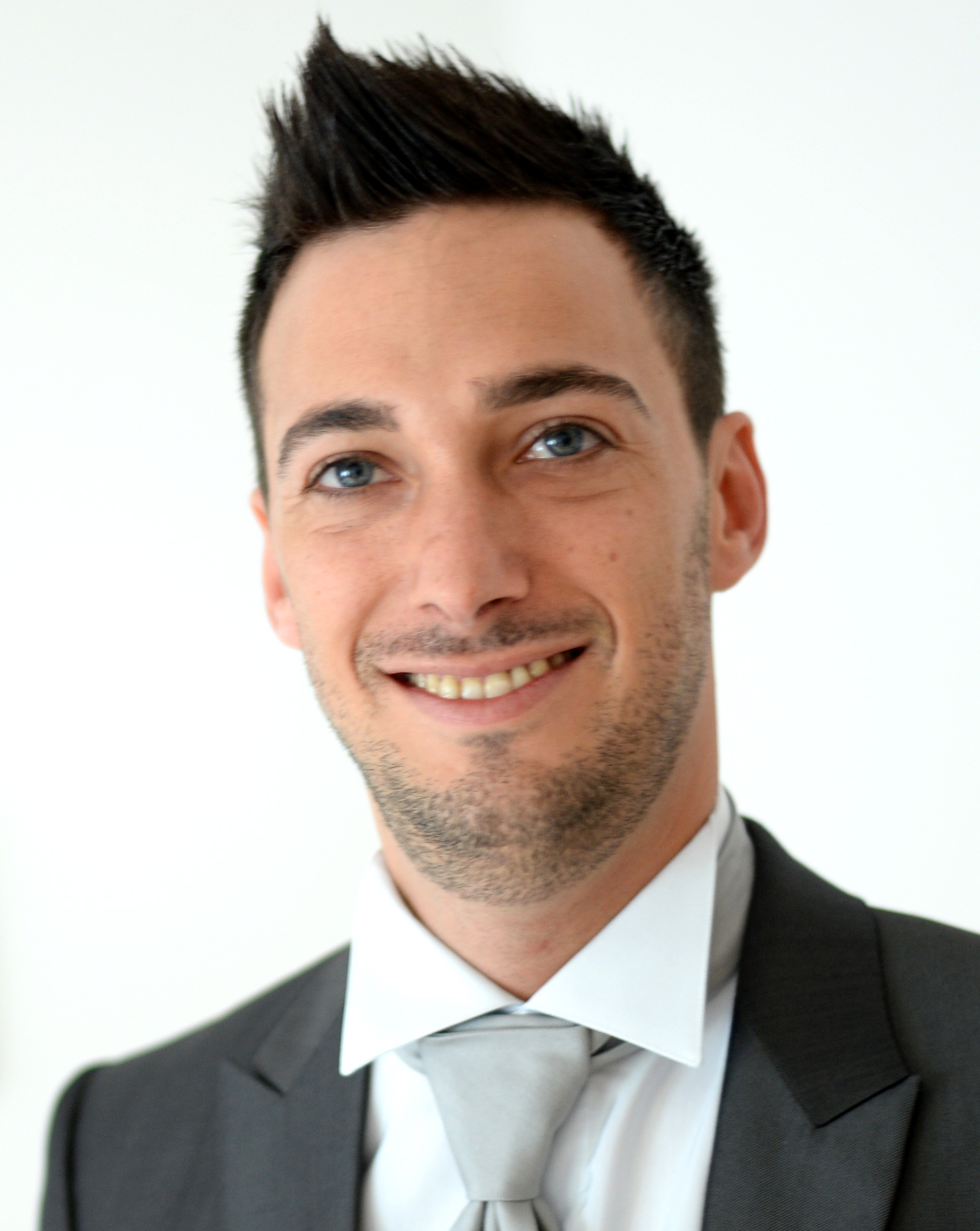}}]{Davide Rossi} (Member, IEEE) received the Ph.D. degree from the University of Bologna, Bologna, Italy, in 2012. He has been a Post-Doctoral Researcher with the Department of Electrical, Electronic and Information Engineering ``Guglielmo Marconi,'' University of Bologna, since 2015, where he is currently an Associate Professor. His research interests focus on energy-efficient digital architectures. In this field, he has published more than 100 papers in international peer-reviewed conferences and journals. He was a recipient of the Donald O. Pederson Best Paper Award 2018, the 2020 IEEE \textsc{Transactions on Circuits and Systems} Darlington Best Paper Award, and the 2020 IEEE \textsc{Transactions on Very Large Scale Integration (VLSI) Systems} Prize Paper Award.
\end{IEEEbiography}

\begin{IEEEbiography}[{\includegraphics[width=0.95in,height=1.25in,clip,keepaspectratio]{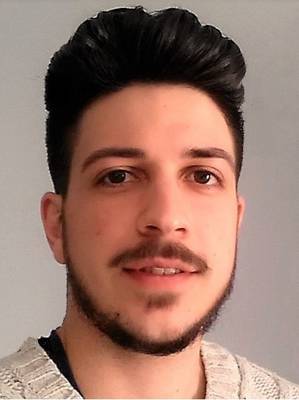}}]{Alfio Di Mauro} (Member, IEEE) received the M.Sc. degree in electronic engineering from the Electronics and Telecommunications Department (DET), Politecnico di Torino, in 2016, and the Ph.D. degree with the Integrated System Laboratory (IIS), Swiss Federal Institute of Technology, Zürich, in 2021.
His research focuses on the design of digital ultra-low power (ULP) system-on-chip (SoC) for event-driven edge computing.
\end{IEEEbiography}

\begin{IEEEbiography}[{\includegraphics[width=0.95in,height=1.25in,clip,keepaspectratio]{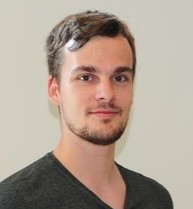}}]{Manuel Eggimann} (Member, IEEE) received the M.Sc. degree in electrical engineering and information technology from ETH Z\"urich, Z\"urich, Switzerland, in 2018, where he is currently pursuing the Ph.D. degree with the ETH Z\"urich Integrated Systems Laboratory.
His research interests include low-power hardware design, edge computing, and very-large-scale integration (VLSI).
Mr. Eggimann was a recipient of the Best Paper Award at the 2019 IEEE 8th International Workshop on Advances in Sensors and Interfaces.
\end{IEEEbiography}

\begin{IEEEbiography}[{\includegraphics[width=0.95in,height=1.25in,clip,keepaspectratio]{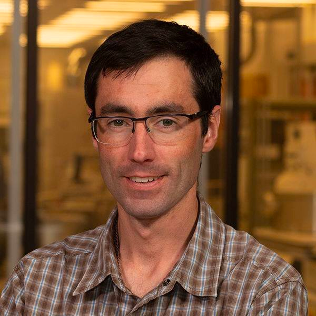}}]{Jorge Tom\'{a}s G\'{o}mez}
Jorge Gómez Mir earned his electrical engineering degree from Universidad de los Andes, Chile, in 2014. Post-graduation, he spent two years as a research engineer at a mining company developing new sensors. In 2016, Jorge started his Ph.D. in emerging semiconductor devices at the Pontificia Universidad Católica de Chile and the University of Notre Dame, USA, completing it in 2021. In September 2021, Jorge started as a Research Scientist at Meta, focusing on emerging semiconductor technologies for AR/VR applications. In January 2024, Jorge moved back to academia as a professor at Universidad de los Andes, Chile.
\end{IEEEbiography}

\begin{IEEEbiography}[{\includegraphics[width=0.95in,height=1.25in,clip,keepaspectratio]{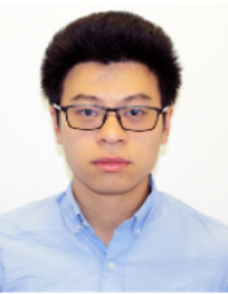}}]{Ziyun Li}
(S’14) received the B.S. degree in electrical and computer engineering from the University of Michigan, Ann Arbor, MI, USA, in 2014, and the Ph.D. degree in electrical engineering from the University of Michigan, Ann Arbor, MI, USA, in 2019.
He is currently with Meta, Redmond, WA, USA. His research interests include SW and HW co-design of high performance, energy efficient AI systems and intelligent cameras to enable next generation intelligent, autonomous systems for AR/VR. Mr. Li was a recipient of the Best Paper Award at the 2016 IEEE Workshop on Signal Processing Systems.
\end{IEEEbiography}

\begin{IEEEbiography}[{\includegraphics[width=0.95in,height=1.25in,clip,keepaspectratio]{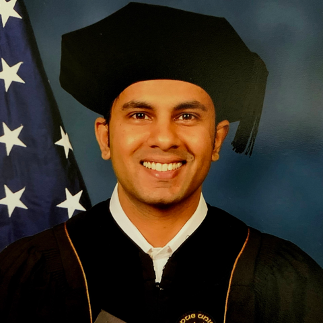}}]{Syed Shakib Sarwar} received the B.Sc. and M.Sc. degrees in electrical and electronic engineering from the Bangladesh University of Engineering and Technology (BUET), Dhaka, Bangladesh, in 2012 and 2014, respectively. In Spring 2019, he completed his Ph.D. degree from Purdue University under the supervision of Prof. Kaushik Roy. Since May 2019, he has been working with Meta Inc. as a Research Scientist. His primary research includes energy efficient algorithms and hardware-software co-design for machine learning applications (deep learning) based on CMOS and emerging devices. His research interest also includes approximate computing in the field of artificial intelligence. At present, his efforts are focused on designing efficient ML systems to mitigate compute constraints for extreme edge contextualized AI use-cases.
\end{IEEEbiography}

\begin{IEEEbiography}[{\includegraphics[width=0.95in,height=1.25in,clip,keepaspectratio]{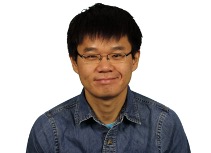}}]{Zhao Wang} earned his Ph.D. in electrical engineering from the University of Texas Dallas, Richardson, TX, in 2012. Since completing his doctoral studies, he has made significant contributions in the semiconductor industry, gaining valuable experience at renowned companies including Qualcomm, Apple, TSMC, and presently, Meta.
Currently serving as a chip lead at Meta, Dr. Wang specializes in low-power chip architecture and development, as well as the exploration of emerging silicon technologies and their applications. His expertise extends to the intricate realm of hardware-software co-design.
Dr. Wang has authored over 10 publications and holds 10 filed patents.
\end{IEEEbiography}

\begin{IEEEbiography}[{\includegraphics[width=0.95in,height=1.25in,clip,keepaspectratio]{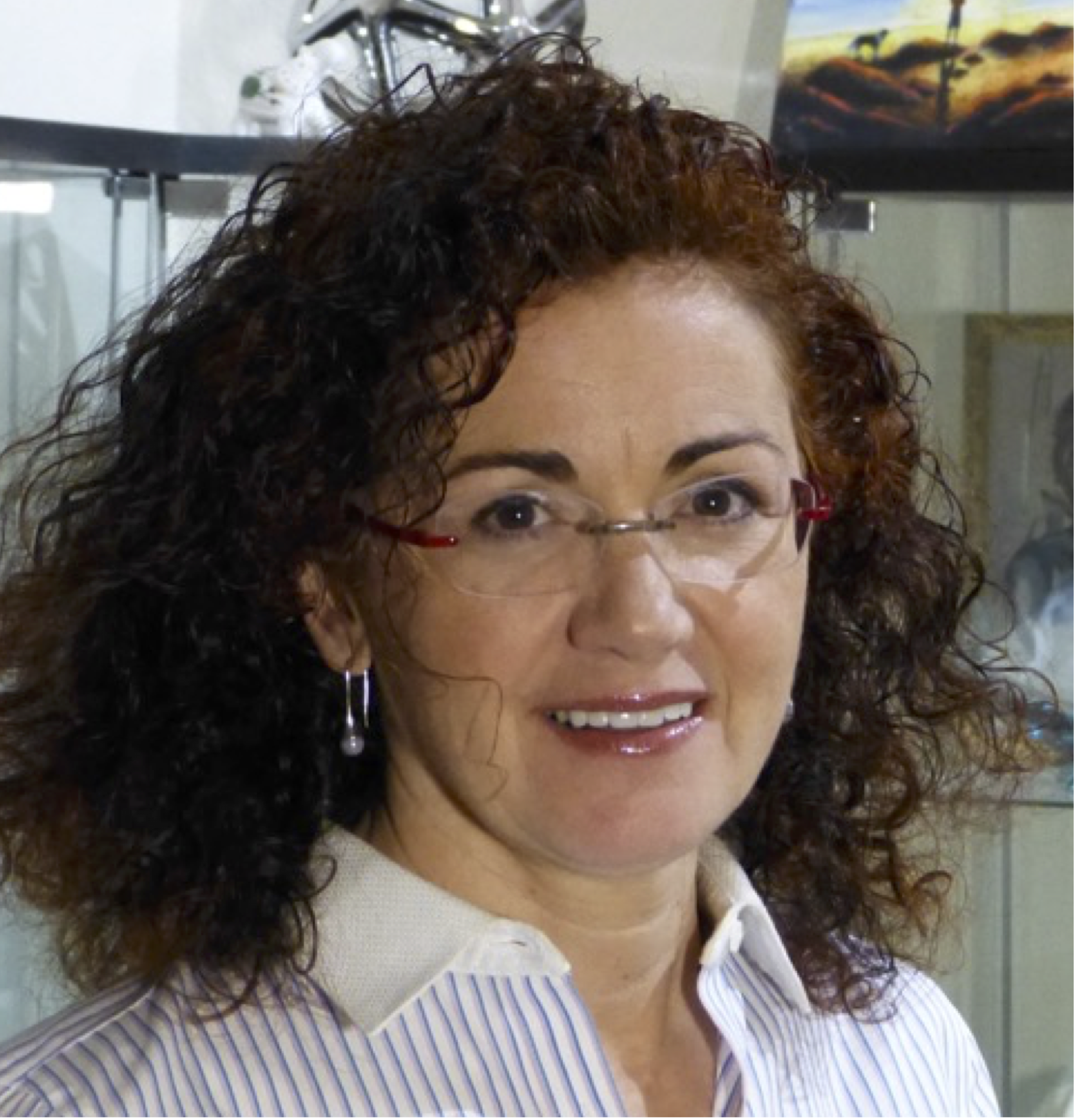}}]{Barbara De Salvo} is Director of Research at Meta Reality Labs Research, responsible for new AI technologies exploration and hardware/software co-design for future Augmented Reality glasses. Before joining Meta in 2019, she was Chief Scientist and Deputy Director of CEA-LETI, driving the path-finding strategy. In 2013-2015, she was manager and visiting scholar in IBM-Albany-NY in the frame of the sub-10nm CMOS International Technology Alliance, where several of her research works have led to product technologies for novel logic ICs (as Silicon-On-Insulator, Finfet and stacked nanowire technology platforms). In CEA-LETI, she founded and led the advanced memory technology division (2008-2013), where she promoted the introduction of disruptive memory technologies, such as phase-change memories, resistive oxide-based and conductive-bridge memories. She pioneered neuromorphic hardware solutions based on emerging technologies for ultra-low-power cognitive systems. She has authored more than 350 referred articles, ten book chapters, a monography on Silicon Non-Volatile Memories edited by Wiley and Sons. She served as General Chair of IEEE IEDM 2022, as well as chair of the IEEE Corporation Award Committee in 2022. She is a Fellow of the IEEE, and an active member of the IEEE Women in Engineering network.
\end{IEEEbiography}

\begin{IEEEbiography}[{\includegraphics[width=0.95in,height=1.25in,clip,keepaspectratio]{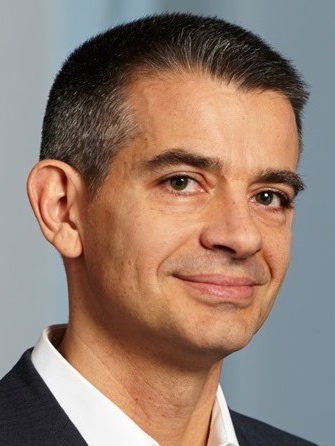}}]{Luca Benini} (Fellow, IEEE) holds the chair of digital Circuits and systems at ETHZ and is Full Professor at the Universit\`a di Bologna. He received a PhD from Stanford University. His research interests are in energy-efficient parallel computing systems and machine learning hardware. He is a Fellow of the ACM and a member of the Academia Europaea. He is the recipient of the 2016 IEEE CAS Mac Van Valkenburg award, the 2020 EDAA achievement Award, the 2020 ACM/IEEE A. Richard Newton Award and the 2023 IEEE CS E.J. McCluskey Award.
\end{IEEEbiography}

\vfill

\end{document}